\documentclass[10pt, preprint2]{aastex}

\usepackage{graphicx}
\usepackage{natbib}
\usepackage{amsmath}
\usepackage{amssymb}
\usepackage{fancyhdr}
\usepackage{latexsym}
\usepackage{color}
\usepackage{txfonts}

\begin{document}

\title{Data-driven Radiative Hydrodynamic Modeling of the 2014 March 29 X1.0 Solar Flare}

\author{Fatima Rubio da Costa\altaffilmark{1}, Lucia Kleint\altaffilmark{2}, Vah\'{e} Petrosian\altaffilmark{1}, Wei Liu\altaffilmark{3} and Joel C. Allred\altaffilmark{4}}
\altaffiltext{1}{Department of Physics, Stanford University, Stanford, CA 94305, USA; Email: frubio@stanford.edu}
\altaffiltext{2}{University of Applied Sciences and Arts Northwestern Switzerland, 5210 Windisch, Switzerland}
\altaffiltext{3}{Bay Area Environmental Research Institute, 625 2nd Street, Suite 209, Petaluma, CA 94952-5159}
\altaffiltext{4}{NASA/Goddard Space Flight Center, Code 671, Greenbelt, MD 20771, USA}

\begin{abstract}
{Spectroscopic observations of solar flares provide critical diagnostics of the physical conditions in the flaring atmosphere. Some key features in observed spectra have not yet been accounted for in existing flare models.}
{Here we report a data-driven simulation of the well-observed X1.0 flare on 2014 March 29 that can reconcile some well-known spectral discrepancies. We analyzed spectra of the flaring region from the {\it Interface Region Imaging Spectrograph} ({\it IRIS}) in \ion{Mg}{2} h\&k, the Interferometric BIdimensional Spectropolarimeter at the Dunn Solar Telescope (DST/IBIS) in H$\alpha$~6563~\AA\ and \ion{Ca}{2} 8542~\AA, and the {\it Reuven Ramaty High Energy Solar Spectroscope Imager} ({\it RHESSI}) in hard X-rays. We constructed a multi-threaded flare loop model and used the electron flux inferred from {\it RHESSI} data as the input to the radiative hydrodynamic code RADYN to simulate the atmospheric response.}
{We then synthesized various chromospheric emission lines and compared them with the {\it IRIS} and IBIS observations. In general, the synthetic intensities agree with the observed ones, especially near the northern footpoint of the flare. The simulated \ion{Mg}{2} line profile has narrower wings than the observed one.}
{This discrepancy can be reduced by using a higher microturbulent velocity (27~km~s$^{-1}$) in a narrow chromospheric layer. In addition, we found that an increase of electron density in the upper chromosphere within a narrow height range of $\approx$~800~km below the transition region can turn the simulated \ion{Mg}{2} line core into emission and thus reproduce the single peaked profile, which is a common feature in all {\it IRIS} flares.}
\end{abstract}


  \section{Introduction}
A large fraction of energy radiated from solar flares originates in the chromosphere. Studying the chromospheric flare emission is therefore a key for understanding how the flare energy is transported and dissipated. 

Chromospheric emission was recorded during the well observed X1.0 flare on 2014 March 29 in the wavelengths H$\alpha$, \ion{Ca}{2} 8542~\AA\ with DST/IBIS \citep{2006SoPh..236..415C} and in \ion{Mg}{2} h\&k with the {\it Interface Region Imaging Spectrograph} \citep[{{\it IRIS}; }][]{2014SoPh..289.2733D}. Since these lines are formed at different heights, the available data cover the whole chromosphere and provide an excellent diagnostic for the flare response of the lower solar atmosphere. However, many of these line profiles form in non-LTE conditions, complicating their interpretation. This requires detailed numerical modeling of the energy input, hydrodynamics, and radiative transfer.

Currently, there are very few non-LTE radiative hydrodynamic codes to model the effects of electron heating on the atmosphere, for example RADYN \citep[see e.g.][]{1997ApJ...481..500C,2005ApJ...630..573A} and FLARIX \citep{2009A&A...499..923K}\footnote{There exist other codes with hydrodynamic response, e.g. \citet{2013ApJ...770...12B}, but they do not include optically thick line emission}.

Past studies using these codes have focused on the atmospheric response to power-law electron beams \citep{2005ApJ...630..573A} and the effects of XEUV backwarming on the \ion{Ca}{2}~H and \ion{He}{1}~10830~\AA\ line profiles \citep{2015ApJ...809..104A}. Comparisons with actual observations are rare. For example,  \citet{2015A&A...578A..72K} used the RADYN code to compare the temporal evolution of the \ion{He}{2} continuum emission with the one observed by EVE MEGS-A during an X1.5 flare or \citet{2015ApJ...813..125K}, who used it to study the evolution of H$\alpha$, but never made a direct comparison with observations.

In our previous paper \citep{2015ApJ...804...56R}, we studied an M3.0 class flare and for the first time made a direct comparison of synthetic line profiles and intensity emission in the chromosphere with observations in H$\alpha$~6563~\AA\ and \ion{Ca}{2}~8542~\AA. Here we take a further step by analyzing a well-observed X1.0 flare, by including more spectral lines, e.g.\,\ion{Mg}{2}, now regularly observed by the {\it IRIS} spacecraft but never simulated for flares, and by making a detailed comparison of spectral line shapes.

Our approach is to use the RADYN code to simulate how accelerated electrons propagate through the solar atmosphere and how the atmosphere responds to the energy deposited by the non-thermal electrons. This allows us to model the dynamic evolution of loops including their temperature and velocity structures and spectral line emission because of the flare. The electron heating is modeled self-consistently from X-ray observations obtained by {\it RHESSI}.

We describe the relevant observations in Section~\ref{Sect:obs} and the simulation setup in Section~\ref{sect:radyn}. The chromospheric emission resulting from our modeling and the comparison with observations are presented in Sections~\ref{Sect:line_profiles} and~\ref{sect:results}. A brief summary and some discussion are presented in Section~\ref{sect:conclusions}.

   \section{Observations}\label{Sect:obs}
\begin{figure}[!htb]
 \centering
 \epsscale{1.03}
 \hspace{-0.3cm}
  \plotone{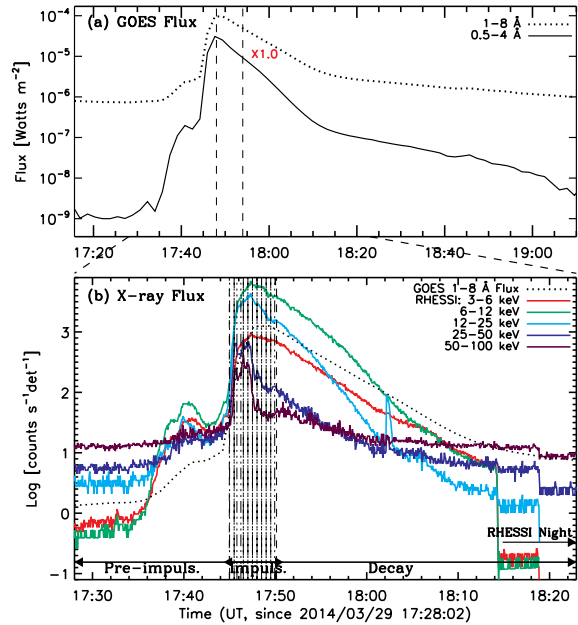}
  \caption{Temporal evolution of the flux measured on 29 March 2014. (a) {\it GOES} soft X-ray flux measured every 3 seconds, showing the X1.0 flare; (b) {\it RHESSI} count rates for different energies in colored solid lines and {\it GOES} 1--8~\AA\ flux. The two vertical dashed lines indicate the duration of the impulsive phase and the vertical dot-dashed lines, the integration time interval of each spectrum.}
  \label{goes_rhessi}
\end{figure}

An X1.0 class flare occurred on March 29, 2014 in active region NOAA~12017 \citep{2015ApJ...806....9K}. It started with a filament eruption at 17:35~UT, leading to the flare, which started at 17:45~UT and reached its maximum at 17:48~UT in the {\it GOES\ }1--8~\AA\ flux. {\it RHESSI} detected X-ray emission from 17:35:28 to 18:14:36~UT, peaking at 17:47:18~UT (see Figure~\ref{goes_rhessi}).

  \subsection{IBIS Observations}\label{sect:ibis}
\begin{figure*}[!htb]
 \centering
 \epsscale{2.15}
  \plotone{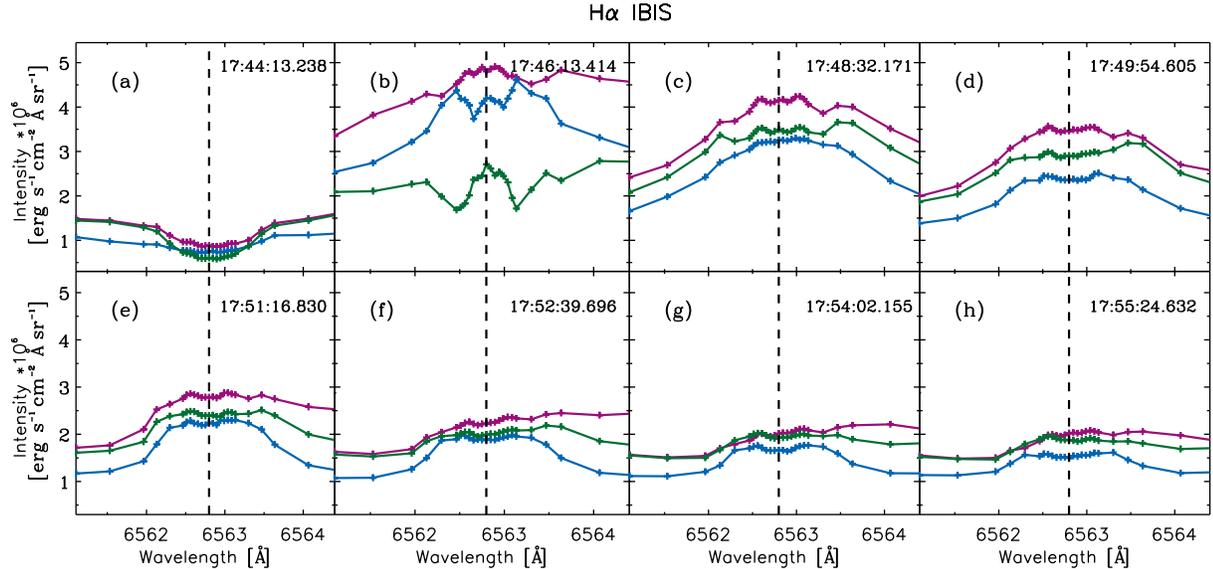}
    \caption{Temporal evolution of the H$\alpha$\ line profile at three different locations: along the flare ribbons (purple for the southern ribbon and blue for the northern ribbon) and just outside the ribbons (green).}
  \label{ha_ibis_evol}
\end{figure*}

IBIS is a full-Stokes polarimeter based on two air-spaced Fabry-Perot interferometers, with a plate scale of 0\farcs1 pixel$^{-1}$ and spectral resolution of 22 and 42~m\AA\ respectively for H$\alpha$\ and \ion{Ca}{2} 8542~\AA\ \citep{2008A&A...481..897R}. Windows of a few Angstrom width around the selected spectral lines are scanned using suitable pre-filters. Six polarization states are reconstructed into images of the four Stokes parameters. We scanned the chromospheric \ion{Ca}{2} 8542~\AA\ and H$\alpha$~6563~\AA\ lines and the photospheric \ion{Fe}{1} 6302~\AA\ line. Here we will focus on the chromospheric emission from H$\alpha$~ and  \ion{Ca}{2} 8542 \AA. The relevant observations were started at 17:30:48~UT and 17:48:08~UT. (see overview in Table~\ref{tabibis}). Their main difference is the decrease of the exposure time from 80~ms to 60~ms and thus their cadence to avoid saturation due to the flare. The IBIS data reduction includes dark and gain corrections, the alignment of all channels, a destretch to correct for the variable seeing, and a correction for the wavelength shift across the field of view due to the collimated Fabry Perot mount and for the prefilter transmission profile. 

H$\alpha$~observations were performed in intensity-only mode (no polarization). Observations from 17:30~UT - 17:47:31~UT were run with 5 repetitions of each of the 25 wavelength points, giving 125 images per sequence. Each sequence took 18~s. From each five identical images, the one with the best seeing was selected, thus assembling image cubes with 25 wavelength points for the analysis. From 17:48:32~UT until the end of the observations, the repetitions of the H$\alpha$~wavelength points were omitted and sequences of 25 images were taken. These sequences took about 4.3~s. Because all spectral lines were scanned sequentially, the overall cadence of the observation is 40-60~s.

\begin{figure*}[!htb]
 \centering
 \epsscale{2.15}
  \plotone{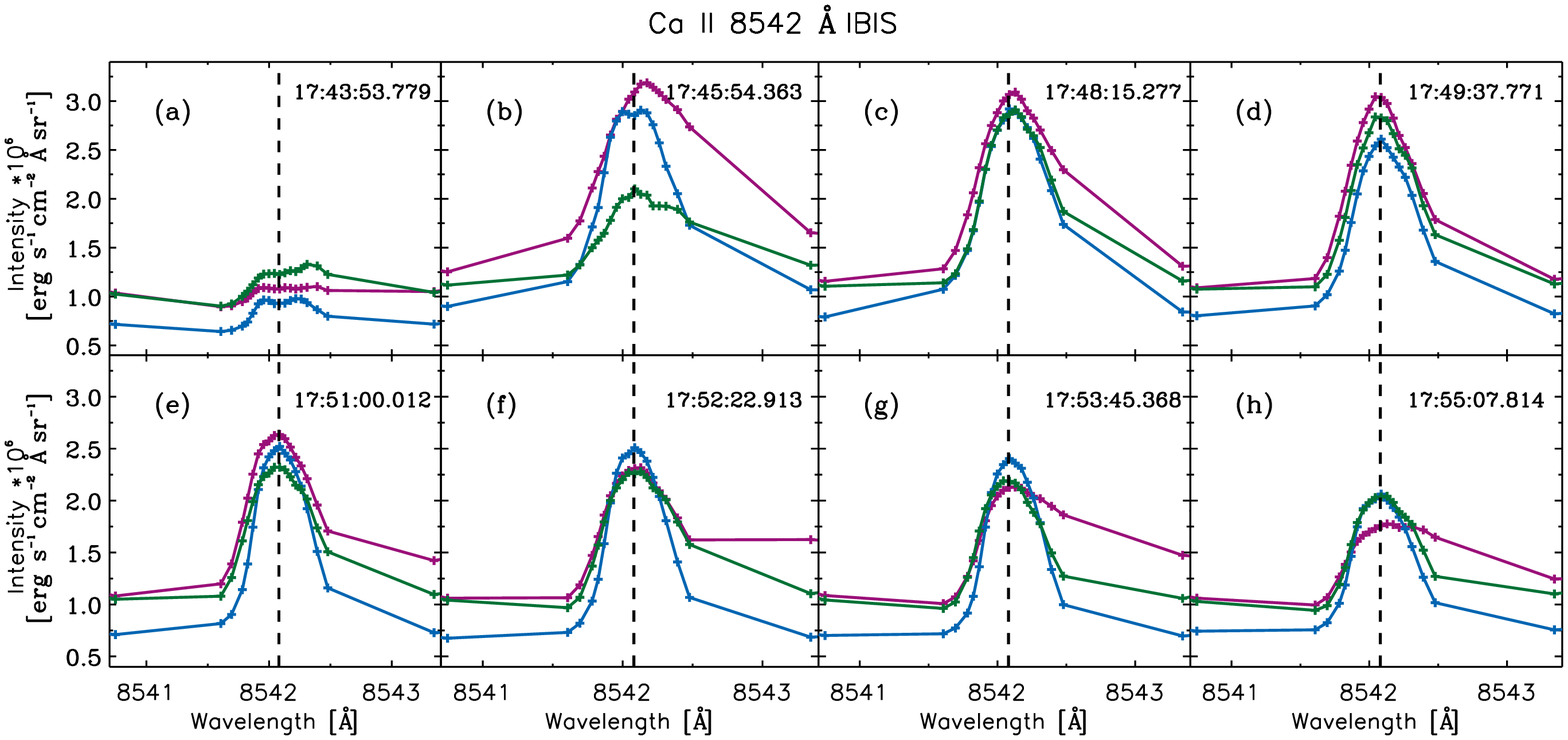}
    \caption{Temporal evolution of the \ion{Ca}{2} 8542~\AA\ line profile at three different locations: along the flare ribbons (purple for the southern ribbon and blue for the northern ribbon) and just outside the ribbons (green).}
  \label{ca_ibis_evol}
\end{figure*}

The \ion{Ca}{2} 8542 \AA\ line was scanned in 21 wavelength points with full-Stokes imaging, giving a total of 126 images per sequence. Depending on the exposure time (cf. Table~\ref{tabibis}), these took 15-18~s for a complete line profile.

\begin{table}[!htb]
   \begin{center}
\caption{Overview of the relevant IBIS observations.}
\setlength\tabcolsep{4pt}
    \begin{tabular}{l | c | c | c}
\tableline
\tableline
\textbf{17:30:48 UT}  & \ion{Ca}{2} 8542 & H$\alpha$\ 6563 & \ion{Fe}{1} 6302\\
\tableline
\# images per seq.  & 126 & 125  & 138 \\
\# wavelengths & 21 & 25 & 23 \\
 time per line  & 18~s  & 18~s  & 20~s \\
\tableline
\textbf{17:48:08 UT}  & \ion{Ca}{2} 8542 & H$\alpha$\ 6563 & \ion{Fe}{1} 6302 \\
\tableline
\#  images per seq. & 126 & 25  & 138  \\
\# wavelengths & 21 & 25 & 23 \\
 time per line & 15.5~s & 4.3~s & 17~s \\
\tableline
\end{tabular}
\end{center}
\label{tabibis}
\end{table}

Figures~\ref{ha_ibis_evol} and \ref{ca_ibis_evol} shows the temporal evolution of the H$\alpha$\ and  \ion{Ca}{2} 8542~\AA\ line profiles respectively, at three different locations: just outside the ribbons (green), over the southern ribbon (purple) and in the northern ribbon (blue) (for exact locations, see Figure~\ref{ha_radyn_ibis}). The intensity calibration was performed comparing the quiet Sun profiles with the ones from RADYN at the same heliocentric angle (see Section~\ref{obs_radyn_ha_ca} for more details).

  \subsection{{\it IRIS} Observations}\label{sect:iris}
  
\begin{figure*}[!htb]
 \centering
 \epsscale{2.03}
  \plotone{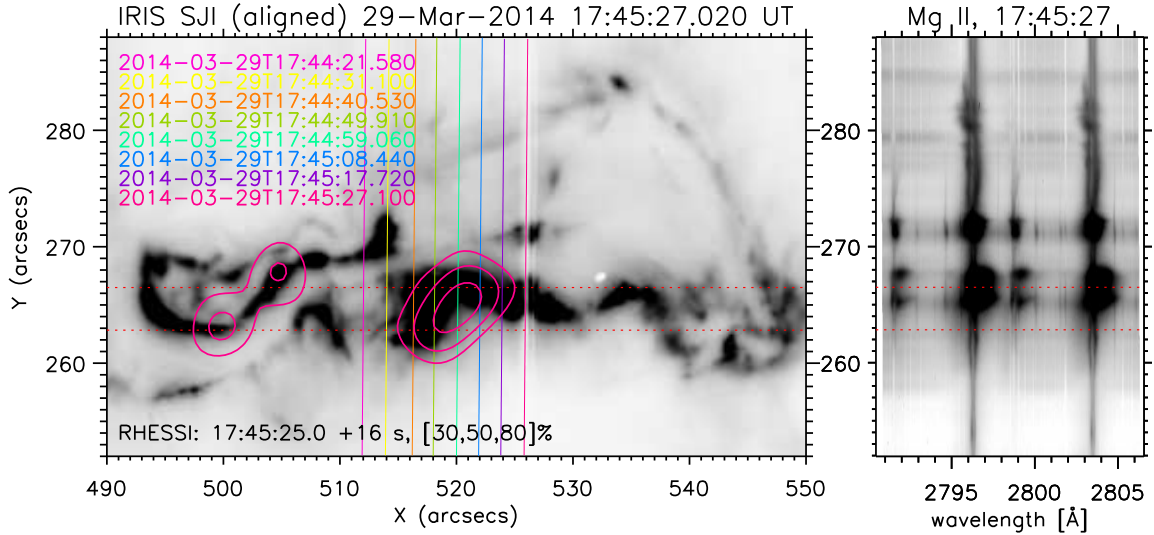}
  \caption{Location of the {\it IRIS} slit at different times (color-coded) on an {\it IRIS}~SJI 2796 background image. The pink contours represent the 30\%, 50\%, and 80\% contours of the {\it RHESSI} HXR emission at 30-70~keV using the CLEAN algorithm for detectors 2, 3 and 4. The right panel shows (color-inverted) \ion{Mg}{2} spectra during the impulsive phase with strong red-asymmetries. The horizontal dotted lines are for reference only to indicate the same positions in both panels.}
  \label{iris_slit}
\end{figure*}

\begin{figure*}
 \centering
 \epsscale{2.2}
  \hspace{-0.3cm}
  \plotone{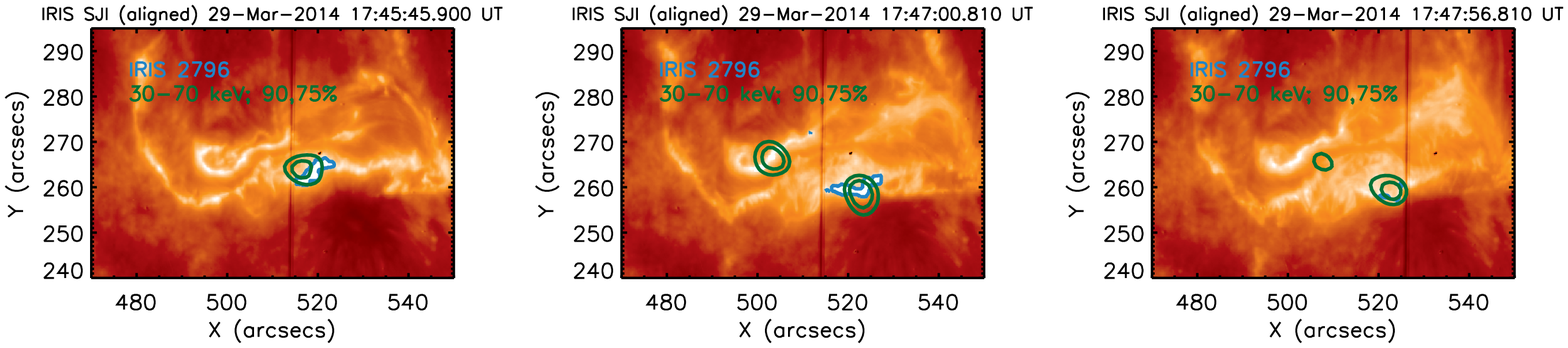}
  \caption{Evolution of the \ion{Mg}{2} 2796~\AA\ {\it IRIS}~SJI during the impulsive phase of the flare. The green contours represent the 90 and 75\% of the {\it RHESSI} 30-70~keV emission and the blue contours display the selected area of the footpoints (above 1212 DN/s).}
  \label{rhessi_footpoints}
\end{figure*}

Coordinated observations by {\it IRIS} were carried out with an eight-step raster of 2\arcsec\ steps, yielding a FOV of 14\arcsec~$\times$~174\arcsec. As shown in Figure~\ref{iris_slit} the {\it IRIS} slit (8 vertical slit positions in different colors) was positioned across the flare ribbons. Complementary slit-jaw images (SJI) were obtained with the filters at 1400~\AA, 2796~\AA, and 2832~\AA. The raster cadence was 75 s, while the exposure time of {\it IRIS}~NUV spectra decreased from 8 s to 2.44 s at 17:46:13.98~UT. {\it IRIS}' plate scale is comparable to IBIS, with 0\farcs167 pixel$^{-1}$. {\it IRIS}' spectral sampling is 0.025~\AA\ pixel$^{-1}$ and covered the \ion{Mg}{2} h\&k lines. IRIS data were converted into absolute units using the official calibration from SolarSoft.

The {\it IRIS}~2796~\AA\ SJI allowed us to estimate the cross-sectional area of the footpoints in the chromosphere. We selected the footpoints area as the emission above a threshold intensity of 1212 DN/s (see blue contours in Figure~\ref{rhessi_footpoints}), obtaining an area of the order of $10^{17}$ cm$^2$. The evolution of the area of the footpoints is shown in Figure~\ref{rhessi_non_thermal_e}(d), where the uncertainty in the area is given by the pixel resolution (0$''$.167) and the temporal error bar is the integration time (8~s).

  \subsection{\textbf{\textit{RHESSI}} Observations}\label{sect:rhessi}
{\it RHESSI} had full coverage of the impulsive phase of the flare (see Figure~\ref{goes_rhessi}). Using the PIXON algorithm, we reconstructed images at 30-70 keV integrated over 20~s intervals from 17:43:00 to 17:49:56 UT. These images revealed two HXR footpoints (see Figure~\ref{rhessi_footpoints}). The southern HXR source moved south-west and crossed slits 3-7 from 17:45:35~UT to 17:46:40~UT \citep{2015arXiv151100480L}.

  \subsubsection{Inferring the non-thermal electron distribution}\label{sect:rhessi_spectra}
To derive the temporal evolution of the flare, we fit the spatially integrated X-ray spectra. The HXR emission from the corona is negligible compared to the footpoints; therefore, we assume that the spatially integrated spectra are dominated by footpoint emission. This provides a reasonable estimate for the electron energy flux as an input to our RADYN simulation.

We analyzed the spectra of the detectors separately and excluded detectors~2, 3, 7, and 8 because of abnormally high threshold and/or low energy resolution \citep{2002SoPh..210...33S} and the higher than average $\chi^2$ values obtained for the spectral fitting. For each of the remaining five detectors, we obtained photon spectra by integrating 30~s intervals from 17:45:00 to 17:49:56~UT, avoiding the change of attenuator from A1 to A3 between 17:45:48 and 17:46:14. The time interval of each spectrum are shown as vertical dotted-dashed lines in Figure~\ref{goes_rhessi}.

\begin{figure}[!htb]
 \centering
 \epsscale{1.03}
  \hspace{-0.3cm}
  \plotone{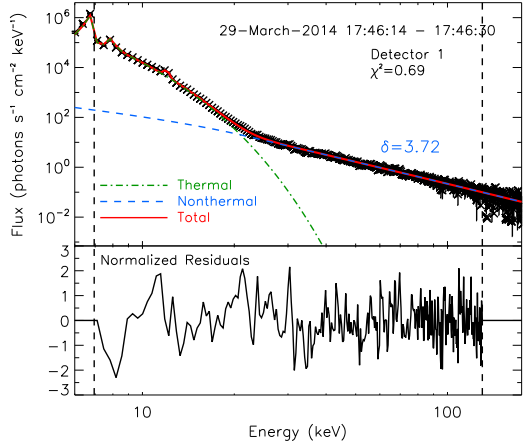}
  \caption{{\it RHESSI} photon spectrum from detector~1, between 17:46:14 and 17:46:30~UT. The green, blue, and red lines show the thermal, non-thermal, and total fitting components. The two vertical dotted lines mark the energy range over which the spectral fit was performed. The bottom panel shows the fit residuals normalized by the one-sigma uncertainties.}
  \label{rhessi_spectrum}
\end{figure}

As \citet{2015ApJ...813..113B} showed, the pileup effect becomes important after $\approx$~17:49:30 UT. Therefore we applied corrections for instrumental emission lines, pulse pileup and the detector response matrix (DRM). The spectra were fitted to a thermal component plus a thick-target, non-thermal component consisting of a broken power law. Figure~\ref{rhessi_spectrum} shows an example of a spectrum from detector 1, taken between 17:46:12 and 17:46:28 UT during the impulsive phase, fitted at low energies to a thermal component and to a non-thermal component at higher energies. 

\begin{figure}[!hbt]
 \centering
 \epsscale{1.03}
  \plotone{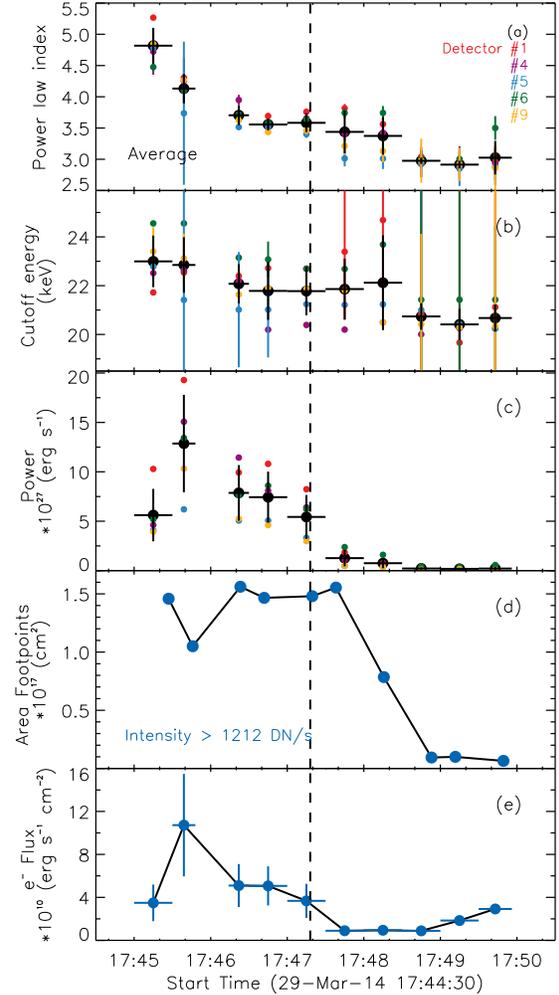}
  \caption{Temporal evolution of the non-thermal electron spectral parameters obtained from fitting the {\it RHESSI} spectra obtained from different detectors (a)-(c); the area of the footpoint emission in 2796~\AA\ (d), (the uncertainty in the area is given by the the pixel resolution (0$''$.167) and the temporal uncertainty is the integration time (8~s)); the non-thermal electron flux (e). The error bar in the y-axis is the standard deviation resulting from all the {\it RHESSI} detectors. The horizontal bars show the integration time of each spectrum. The dashed line indicates the time of the maximum integrated X-ray flux from {\it RHESSI}.}
  \label{rhessi_non_thermal_e}
\end{figure}

Averaging the spectral parameters of the non-thermal electrons for all selected detectors, the resulting power of electrons shows that its maximum is reached before the peak in GOES X-ray flux (Figure~\ref{rhessi_non_thermal_e}). This is because at later times the thermal component contributes more to the total electron flux.

To estimate the electron flux, we divided the power of non-thermal electrons by the cross-sectional area of the footpoints in {\it IRIS} 2796~\AA\ (Figure~\ref{rhessi_non_thermal_e}(d)) interpolated to the times of each spectrum. Figure~\ref{rhessi_non_thermal_e}(e) shows the temporal evolution of the flux of non-thermal electrons, which peaks at 17:45:39~UT, agreeing with the flux estimation of \citet{2015ApJ...813..113B}. The flux error bars have been calculated as the standard deviation resulting from all the {\it RHESSI} detectors and the temporal error bar show the integration time of each spectrum. The average cutoff energy and spectral index in Figure~\ref{rhessi_non_thermal_e}(a) and Figure~\ref{rhessi_non_thermal_e}(b), together with the electron energy flux in Figure~\ref{rhessi_non_thermal_e}(e) are the input parameters for our RADYN simulation described next.

  \section{RADYN Simulations}\label{sect:radyn}
We used the RADYN code of \citet{1997ApJ...481..500C}, including the modifications of \citet{1999ApJ...521..906A} and \citet{2005ApJ...630..573A}, to simulate the radiative-hydrodynamic response of the lower atmosphere to energy deposition by non-thermal electrons in a flare loop, following the same description as in \citet{2015ApJ...804...56R}. The line transitions treated in detail are listed in Table~1 of \citet{1999ApJ...521..906A}. 

  \subsection{Simulation Setup}\label{Sect:radyn_setup}
We assumed a single quarter circle loop geometry of 10~Mm height with a diameter of $4 \times 10^8$ cm$^2$ as determined from the {\it IRIS\ } observations, in a plane-parallel model atmosphere. As a boundary condition, we assumed reflection at the loop apex and an open boundary at the bottom of the loop.

\begin{figure}[!htb]
 \centering
 \epsscale{1.03}
  \plotone{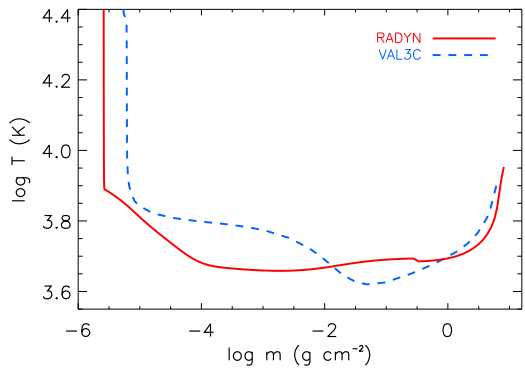}
  \caption{Logarithm of the temperature as function of column mass for the modified pre-flare atmosphere used in this paper (red) and the semiempirial VALC model from \citet{1981ApJS...45..635V} (blue dashed).}
  \label{preflare_atm}
\end{figure}

The initial atmosphere (red line in Figure~\ref{preflare_atm}) is a modification of the FP2 model of \citet{1999ApJ...521..906A}, where the temperature was fixed at 10$^6$~K at the loop top and decreases at the bottom of the loop, such that the photospheric temperature closely fits the one of the Sun \citep[5780~K; ][]{fraknoi2000voyages}. The atmosphere stands in hydrodynamic equilibrium state, resulting from relaxing it without providing any external heating.

Our modified atmosphere has a lower temperature, density and pressure in the chromosphere than the FP2 model of \citet{1999ApJ...521..906A} and is cooler and has lower electron density than the VALC quiescent state from \citet{1981ApJS...45..635V}. The continuum emission fits to a Planck function with a photospheric temperature of 5806~K, optimally fitting the pre-flare continuum emission observed at four wavelength bands from the UV to the IR: at 1333~\AA\ ({\it IRIS}~FUV), 2825~\AA\ ({\it IRIS}~NUV), 6173~\AA\ (SDO HMI), and 10840~\AA\ (DST/FIRS) \citep[see][for more details]{2016ApJ...816...88K}. The continuum in the {\it IRIS}~FUV and NUV is formed at lower temperatures (higher in the atmosphere), and therefore these points are expected to lie below the fitted Planck function.

The non-thermal electron heating estimated from {\it RHESSI} (see Section~\ref{sect:rhessi_spectra}) was included in RADYN as a source of external heating in the equation of internal energy conservation \citep[for more details see][]{2015ApJ...804...56R}.
 
 \subsection{Multi-thread setup}\label{sect:multiloops}
The impulsive phase of the flare lasts about four minutes, but it is unlikely that flare-accelerated particles precipitate down in a single flux loop for the duration of the entire impulsive phase \citep{2006ApJ...637..522W, 2010ApJ...725..319Q}. Moreover, the footpoints move about 15 arcsec south-west during the flare, so the assumption of a single thread loop would not be a good representation of this particular flare.

\citet{2012arXiv1202.4819H} applied the idea of multi-thread simulations by describing a number of free parameters, adjusted to match the observed coronal emission. \citet{2010SoPh..267..107L} constrained the heating rate and duration of each thread using a reconnection model and comparing the estimated coronal radiation with observations. \citet{2012ApJ...752..124Q} employed multi-threads and estimated the flare heating from ultraviolet light curves. Moreover, as \citet{2015ApJ...813...70F} have argued, applying a continuous heating flux variable in time along a single loop would be considered the debatable extreme case of a multithread structure of loops with a filling factor of 1.

Here we adopted the idea of multithread structure of loops by running the RADYN code at different times applying the heating rate estimated from the {\it RHESSI} spectra as explained in Section~\ref{sect:rhessi_spectra}. The essential difference between multi-thread and a uniform loop is that each new simulation, representing a new thread, starts from the initial atmosphere of Figure~\ref{preflare_atm} and initial conditions explained in Section~\ref{Sect:radyn_setup}. 

\begin{figure}[!htb]
 \centering
 \epsscale{1.03}
  \plotone{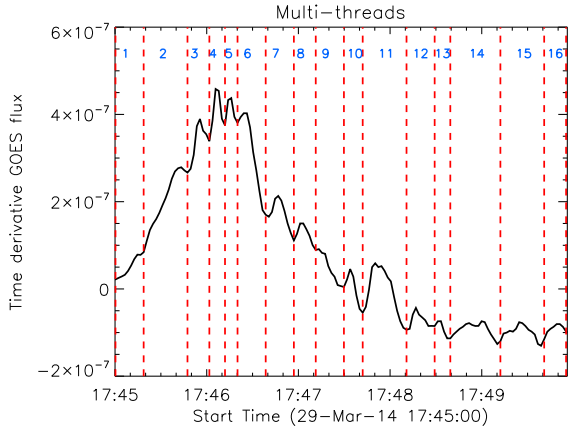}
  \caption{Time derivative of the GOES 1-8~\AA\ light curve. The red dashed lines indicate the beginning and end of each individual thread.}
  \label{goes_deriv}
\end{figure}

As \citet{2006ApJ...637..522W} discussed, the time scale of each individual thread would be of the order of 60 to 80 seconds. These pulses are in fact seen in other instruments \citep[see e.g.][]{2007ApJ...656..577A, 2014ApJ...789...71F}, indicative of distinct heating events. 
In this paper we assume that the spikes shown in the time derivative of the GOES 1-8~\AA\ light curve (Figure~\ref{goes_deriv}) correspond to single bursts. This idea has been previously investigated by \citet{2004ApJ...611L..49W}, based on the assumption that there is a linear correlation between the peak in the time derivative of the soft X-rays and the non-thermal emission, i.e. the so called Neupert effect \citep{1968ApJ...153L..59N}.

By assuming that each burst is due to a new thread, we have estimated that this flare is composed of 16 different threads starting at the times indicated by the vertical dashed lines of Figure~\ref{goes_deriv}. The duration of each spike in the GOES time derivative defines the heating phase of each thread; three times this duration is assumed to be the relaxation phase when electron heating is turned off but the atmosphere is allowed to relax. We use these time scales as a proxy for the lifetime of the heating phase on individual threads. The assumption that each spike is a single burst seems to be a reasonable approximation and agrees with the time scales found by \citet{2006ApJ...637..522W}. One could also use {\it RHESSI} hard X-ray light curves to estimate the duration of one thread. The main spikes in the GOES derivative are temporally coaligned with those in the 25-50~keV flux, which justifies our usage of the GOES derivative.

Considering that the flux of electrons estimated at each time interval in Section~\ref{sect:rhessi_spectra} (see Figure~\ref{rhessi_non_thermal_e}(e)) is the total flux of the whole footpoint, the flux of the heating phase of each single thread was calculated by multiplying the flux of electrons by the trend in the GOES time derivative, such that the total flux of the threads in a given time interval agrees with the total flux for the same interval estimated from the {\it RHESSI} spectra. 

\begin{table}[!htb]
\begin{center}
\caption{Starting time of each individual thread and its time duration during the heating and relaxing phase.}
\begin{tabular}{c c c c}
\tableline
\tableline
Thread \# & Start Time & Duration  & Duration  \\
 & UT & heat. phase & relax. phase \\
\tableline
 1 & 17:45:00 & 18.90~s & 56.70~s \\
 2 & 17:45:19 & 29.40~s & 88.20~s \\
 3 & 17:45:48 & 14.70~s & 44.10~s \\
 4 & 17:46:03 & 10.50~s & 31.50~s \\
 5 & 17:46:13 & 8.40~s & 25.20~s \\
 6 & 17:46:21 & 18.90~s & 56.70~s \\
 7 & 17:46:40 & 18.90~s & 56.70~s \\ 
 8 & 17:46:59 & 14.70~s & 44.10~s \\
 9 & 17:47:13 & 18.90~s & 56.70~s \\
 10 & 17:47:31 & 12.60~s & 37.80~s\\
 11 & 17:47:44 & 29.40~s & 88.20~s\\
 12 & 17:48:13 & 16.80~s & 50.40~s \\
 13 & 17:48:31 & 10.50~s & 31.00~s \\
 14 & 17:48:41 & 33.60~s & 100.80~s \\
 15 & 17:49:14 & 29.40~s & 88.20~s\\
 16 & 17:49:43 & 16.80~s & 50.40~s\\
\tableline
\end{tabular}
\end{center}
\label{table_time_multiloops}
\end{table}

Table~\ref{table_time_multiloops} shows the duration of the heating and relaxing phase of each single thread. The temporal evolution of each simulation agrees with the timescales previously reported by \citet{2015ApJ...811....7L} in the dynamics of the flare ribbons.

The total contribution to the flare is the result of temporally adding up all the sequentially heated threads magnetically confined within our simulated 1D flare loop model. In this way we construct a multi-thread model of the Sun's atmospheric response and emission produced during the flare, as in the spatial evolution of a flare arcade.

  \section{Synthetic chromospheric emission}\label{Sect:line_profiles}
In this section we will present the chromospheric emission in H$\alpha$~and  \ion{Ca}{2}~8542~\AA, resulting from the RADYN simulations (Section~\ref{Sect:ha_radyn}) and in \ion{Mg}{2}~h\&k, resulting from the RH code (Section~\ref{sect:MgII}). The results will be compared to observations in Section~\ref{sect:results}.
 
We follow the same procedure for comparing the H$\alpha$~emission with synthetic RADYN line profiles as \citet{2015ApJ...804...56R}. As explained in their paper, centrally reversed profiles are a consequence of the combination of several effects: the behavior of the temperature stratification at the height where the line core is formed, together with the sudden change in the response of the source function within a very thin atmospheric layer. The comparison with observations was done for the available data limited to a particular time step during the gradual phase of the flare. Here we study the temporal evolution of the line profile during the whole duration of the flare, and as discussed in Section~\ref{sect:multiloops}, we include sequentially heated threads instead of a single loop continuously heated.

\begin{figure*}[htb]
 \centering
 \epsscale{2.1}
  \plotone{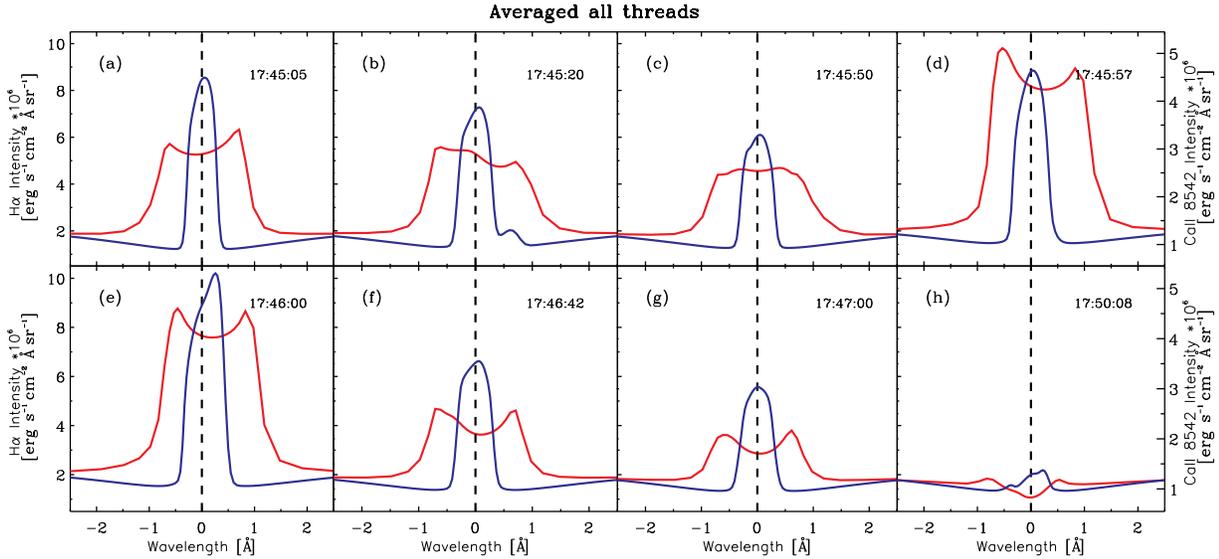}
  \caption{Temporal evolution of the synthetic H$\alpha$~(red) and \ion{Ca}{2} 8542~\AA~(blue) line profiles. The dashed line indicates the center of the lines. Note: the intensity scales for H$\alpha$~and \ion{Ca}{2} 8542~\AA~are different.}
  \label{ha_ca_radyn}
\end{figure*}

We calculate the intensity for each individual thread and, considering that all the threads equally contribute to the total intensity, we temporally averaged the intensity at each wavelength position for all the threads contributing at a specific time interval. 

  \subsection{Formation and Evolution of the H$\alpha$~and  \ion{Ca}{2}~8542~\AA\ Spectral Lines}\label{Sect:ha_radyn}
Focusing first on the H$\alpha$~emission, Figure~\ref{ha_ca_radyn} shows the temporal evolution of the averaged line profile (red profiles). During the impulsive phase, the synthetic H$\alpha$~profile is in emission, often showing a central reversal. The reversal results from a sudden change of the source function behavior at the height where the line core is formed. After 52 seconds, the line profile shows an asymmetry: the blue shifted peak of the line becomes stronger and the line is shifted to higher wavelengths (Figure~\ref{ha_ca_radyn}(d)). Closer to the peak of the flare the red wing of the line becomes stronger and the line center shows a shift towards lower wavelengths (Figure~\ref{ha_ca_radyn}(e)). 

The \ion{Ca}{2} 8542~\AA\ line profile (blue profiles of Figure~\ref{ha_ca_radyn}) responds similarly to plasma motions. During the impulsive phase, the synthetic \ion{Ca}{2} 8542~\AA~profile is in emission and is more sensitive to plasma velocities than H$\alpha$. After 20 seconds, the line shows a secondary small peak at high wavelengths, due to downflow velocities of the order of 30~km~s$^{-1}$ from the contributing threads 1 and 2 in the upper chromosphere, below the region where the core of the line is formed. At later times, closer to the flare peak (Figure~\ref{ha_ca_radyn}(e)) the core of the line shows asymmetries on the red side due to downflow velocities in the upper chromosphere, exactly in the core formation region. In the gradual phase of the flare (Figure~\ref{ha_ca_radyn}(h)), the line profile is fainter, but still above the continuum emission.

\citet{1981ApJS...45..635V} \citep[and recently][in flares]{2015ApJ...813..125K} found that the \ion{Ca}{2}~8542~\AA\ line is formed deeper in the atmosphere than H$\alpha$. Moreover, their simulated profile shows a weak downflow at the height of core formation which shifts the line center to the red, agreeing with our profiles close to the peak of the flare.

\begin{figure*}[!hbt]
 \centering
 \epsscale{1.05}
  \mbox{
    \mbox{\includegraphics[scale=0.4]{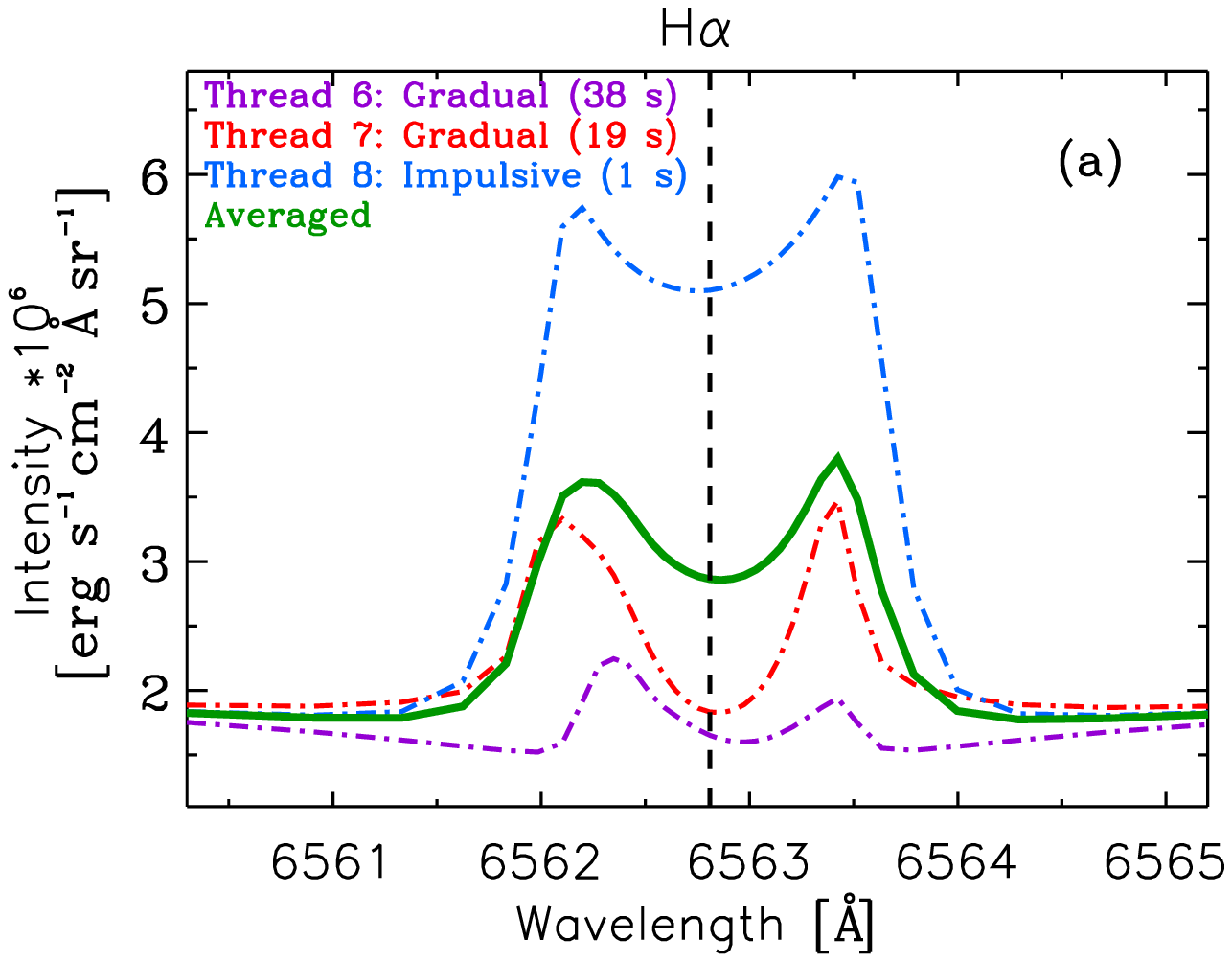}}   
    \hspace{-0.3cm}
    \mbox{\includegraphics[scale=0.4]{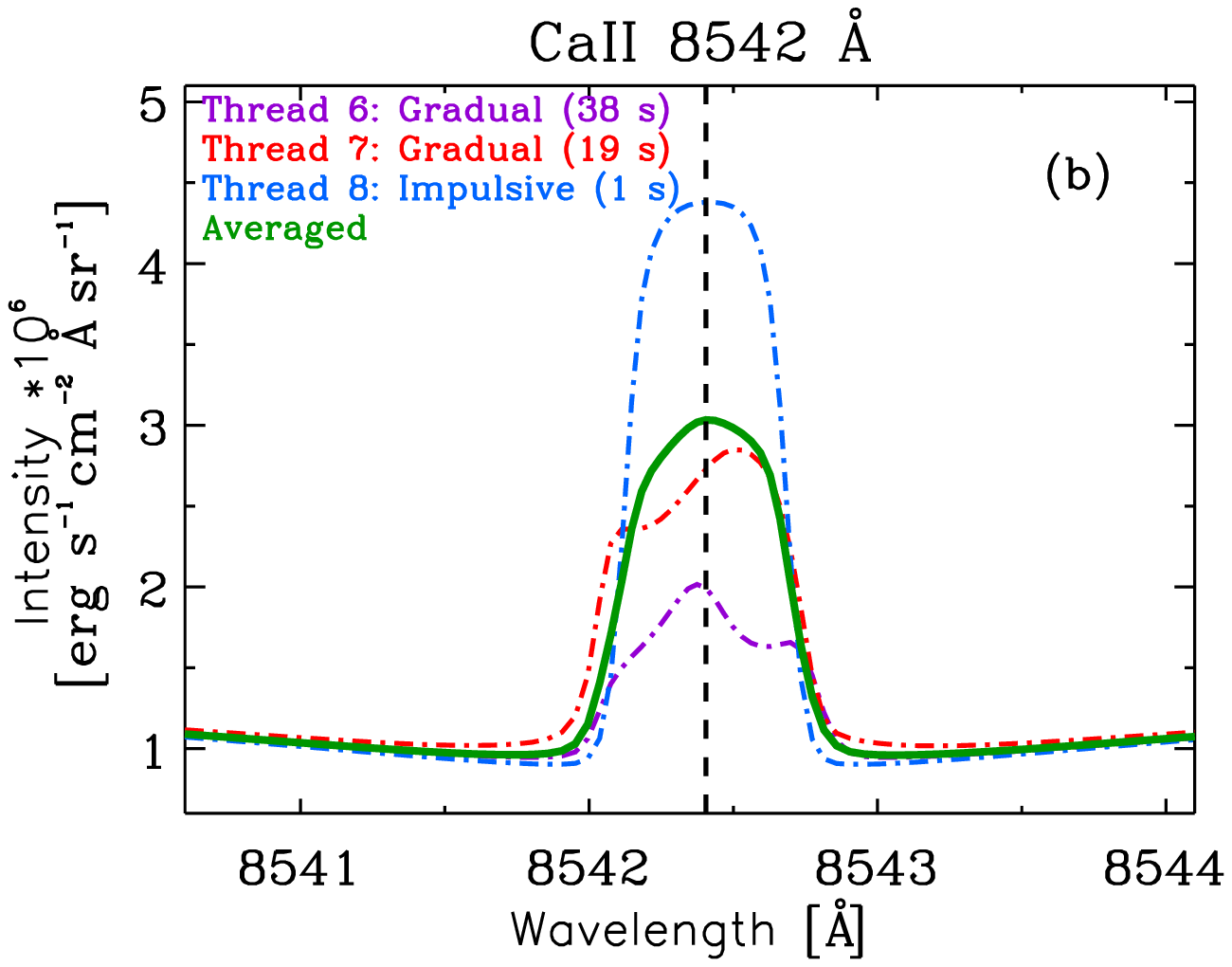}}
    \hspace{-0.3cm}
    \mbox{\includegraphics[scale=0.4]{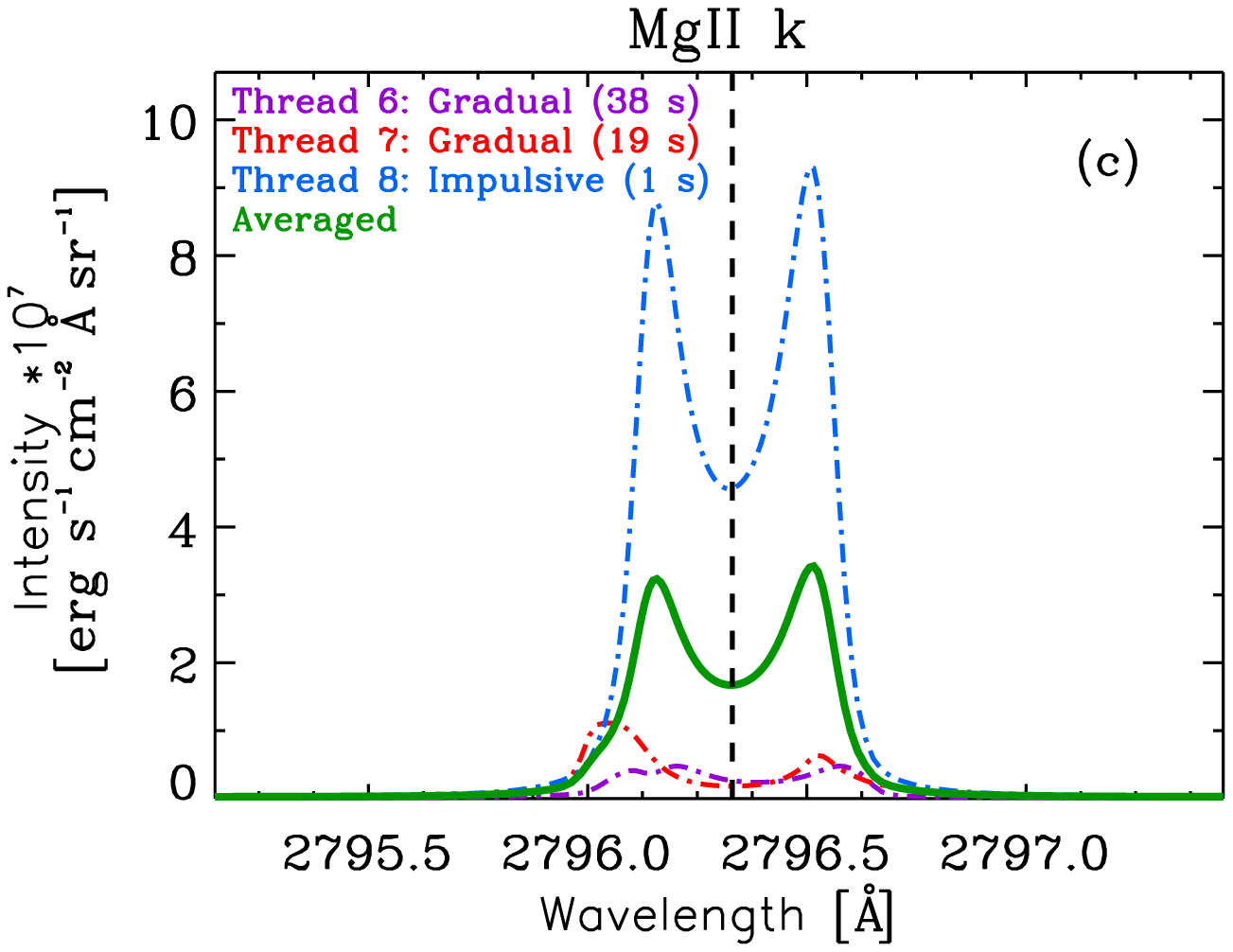}}
  }
    \caption{H$\alpha$~line, \ion{Ca}{2} 8542~\AA\ profiles (from RADYN) and \ion{Mg}{2}~k profiles (from RH) at 17:47:00 UT (green) and the contribution of threads~6 (dotted-dashed purple),~7 (dotted-dashed red) and~8 (dotted-dashed blue) separately.}
    \label{ha_ca_mg_t120_loops}
\end{figure*}

We selected a specific time interval (17:47:00 UT) to analyze how the emission of each single thread contributes to the total intensity at each wavelength range. As Table~\ref{table_contribution_loops} shows, at 17:47:00~UT the total intensity is formed by the contribution of the threads~6 and 7 in their relaxing phase (after 38 and 19 seconds of simulation) and thread~8 in its heating phase (after 1~second) with a non-thermal electron heating of $3.6\times 10^{10}$ erg s$^{-1}$ cm$^{-2}$. The line profiles of each single contributing thread and the averaged line profile are shown in Figure~\ref{ha_ca_mg_t120_loops} for H$\alpha$\ (panel~a),  \ion{Ca}{2}~8542~\AA\ (panel~b), and \ion{Mg}{2} k (panel~c).

\begin{figure*}[!hbt]
 \centering
 \epsscale{2.15}
  \plotone{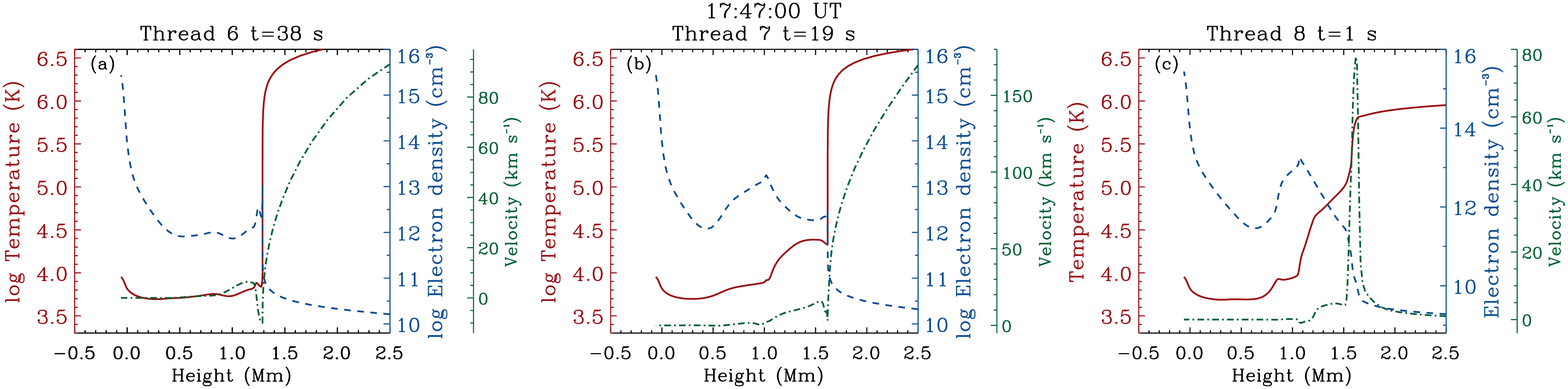}
  \caption{Chromospheric stratification of the temperature (red), electron density (blue), and velocity (green) for the three different threads contributing to the total flaring emission at 17:47:00 UT. Positive velocities indicate plasma moving upwards, towards the corona, and corresponds to blueshifts in the line profile.}
  \label{chromosphere_loops6_7_8}
\end{figure*}

Asymmetries generally result from velocities, especially downflows during the flare maximum. As an example, the line profile resulting from thread~6 (blue line) in Figure~\ref{ha_ca_mg_t120_loops} shows a stronger red peak due to downflows of almost 15~km~s$^{-1}$ (see right panel in Figure~\ref{chromosphere_loops6_7_8}). In the gradual phase, the H$\alpha$~profile returns to absorption. Table~\ref{table_contribution_loops} shows which threads contribute to the averaged line profile (green line in Fig.~\ref{ha_ca_mg_t120_loops}) at the selected time steps. The evolution times of the individual threads are given in the legend.

\citet{2015ApJ...813..125K} studied the H$\alpha$~emission during an M1.1 class flare and stressed that the red and blue shifts visible in the line profile may not be associated with plasma down- and upflows. They also found that the core of the line remains unshifted before the flare and shows redshifts after the flare maximum, agreeing with the behavior of our synthetic H$\alpha$~profiles.

\begin{table}[!htb]
\begin{center}
\caption{Contribution of the different threads at each time step shown in the temporal evolution of the different wavelengths shown in Figures~\ref{ha_ca_radyn} and \ref{mg_loops}.}
\begin{tabular}{c c c}
\tableline
\tableline
Time (UT) & Contributing Threads & Time-step \\
& & of each thread\\
\tableline
17:45:05 UT & Thread 1 & Heating (5~s)\\
17:45:20 UT  & Thread 1 & Relaxing (20~s)\\
 & Thread 2 & Heating (2~s)\\
17:45:50 UT  & Thread 1 & Relaxing (50~s)\\
 & Thread 2 & Relaxing (31~s)\\
 & Thread 3 & Heating (2~s)\\
17:45:57 UT & Thread 2 & Relaxing (38~s)\\
 & Thread 3 & Relaxing (9~s)\\
17:46:00 UT & Thread 2 & Relaxing (41~s)\\
 & Thread 3 & Heating (12~s)\\
17:46:42 UT  & Thread 6 & Relaxing (20~s)\\
 & Thread 7 & Heating (1~s)\\
17:47:00 UT  & Thread 6 & Relaxing (38~s)\\
 & Thread 7 & Relaxing (19~s)\\
 & Thread 8 & Heating (1~s)\\
17:50:08 UT & Thread 14 & Relaxing (85~s)\\
 & Thread 15 & Relaxing (52~s)\\
 & Thread 16 & Relaxing (22~s)\\
\tableline
\end{tabular}
\end{center}
\label{table_contribution_loops}
\end{table}

The H$\alpha$~line profile of thread~6 shows a reversal in the center due to the sudden change of the source function at 1.3~Mm, where the photons of the line core are formed. This is due to the temperature stratification at this height (i.e. right below the beginning of the transition region). Similarly, in thread~7, this change occurs at 1.1~Mm; where the source function separates from the Planck function, changing the direction drastically with frequencies.

In general, the asymmetry in the core of both H$\alpha$~and \ion{Ca}{2}~8542~\AA\ profiles, is the results of the velocity distribution at the formation height of the line core (see the velocity distribution in Figure~\ref{chromosphere_loops6_7_8}). 

The \ion{Ca}{2}~8542~\AA\ line profile resulting from threads~6 and 7, both show an asymmetry in the line core due to downflow velocities in the region where the line core is formed (which is very similar to the one of H$\alpha$); while the line profile resulting from thread~8 during the beginning of its impulsive phase shows a symmetric profile because the photons at the core of the line are formed at a lower height than in H$\alpha$~($\approx$1.05~Mm); therefore the downflow velocities do not affect the line profile. In general, the photons formed in the wings of the \ion{Ca}{2}~8542~\AA\ line profile are formed in a layer above those of H$\alpha$; while the core formation of both lines is located roughly at the same height.

To study how the atmosphere behaves in the upper chromosphere, Figure~\ref{chromosphere_loops6_7_8} shows the stratification in height of the temperature (red), electron density (blue) and velocity (green) for the different threads contributing to the total flaring emission at 17:47:00 UT: thread~6 after 38 seconds (left), thread~7 after 19 seconds (middle) and thread~8 after 1 second (right). The threads~6 and 7 are already in their relaxing phase; while in thread~8 the electrons are starting to interact with the atmosphere.

The temperature in the corona of thread~6 is decreasing; the pressure work has pushed the transition region towards lower layers (1.28~Mm) with respect to the initial times (1.57~Mm). The atmosphere of thread~7 is starting its gradual phase; the temperature in the corona is decreasing, the electron density is stable in the corona and decreasing in the chromosphere. The atmosphere resulting from thread~8 is at the beginning of its impulsive phase, therefore the atmosphere is very similar to the one in HD equilibrium.

  \subsection{Formation and Evolution of the \ion{Mg}{2} h\&k Spectral Lines}\label{sect:MgII}
 As \citet{2013ApJ...772...89L} pointed out, due to the low chromospheric density, the time average between collisions is larger than the lifetime of an \ion{Mg}{2} ion in the upper levels of the lines. For this reason, the frequency of emitted and absorbed photons in scattering processes is correlated and the assumption of complete frequency redistribution (CRD) is not valid. Therefore, the \ion{Mg}{2} h\&k lines in the quiet Sun and plage are strongly affected by the effect of partial frequency redistribution (PRD). Since RADYN calculates the emission of the \ion{Mg}{2} lines assuming CRD, we employ the RH code \citep{2015A&A...574A...3P, 2001ApJ...557..389U} to calculate the \ion{Mg}{2} emission with PRD. We perform non-LTE radiative transfer computations with a modified version of the RH code, which includes the heating due to the injection of electrons in the atmosphere. This code treats the effects of angle-dependent PRD using the fast approximation by \citet{2012A&A...543A.109L}.

\begin{figure*}[!htb]
 \centering
 \epsscale{2.05}
  \plotone{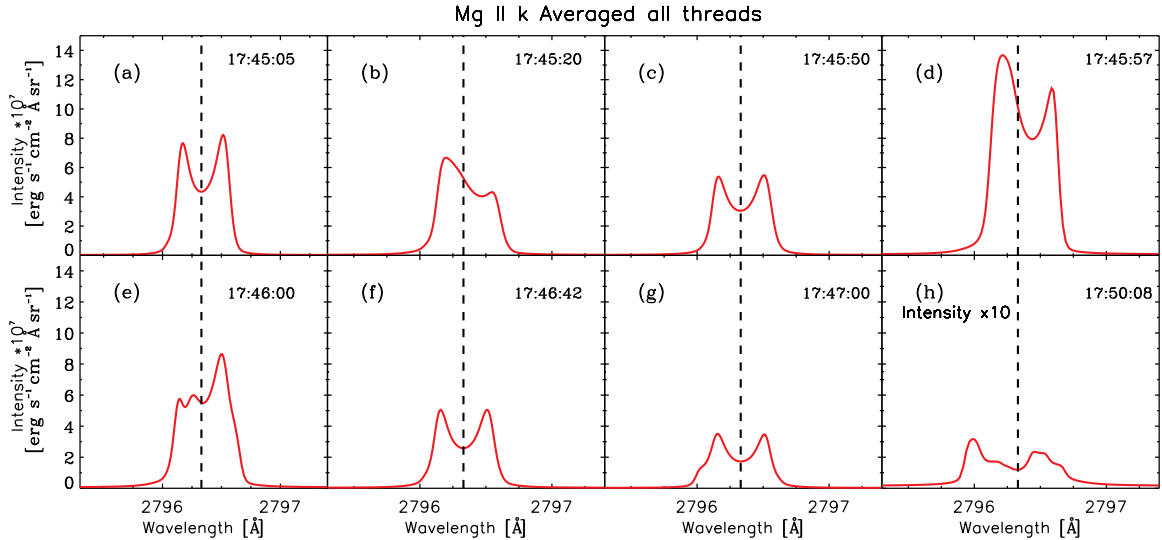}
  \caption{Evolution of the synthetic \ion{Mg}{2}~k~line profile calculated with the RH code. The dashed line represents the line center.}
  \label{mg_loops}
\end{figure*}

We considered a time series of 1D hydrodynamical snapshots computed with RADYN with a timescale of 1~second, which is shorter than the timescale for ionization/recombination and used a 10 level plus continuum model atom for \ion{Mg}{2} as described in \citet{2013ApJ...772...89L}. We compared the RH calculation with PRD to the RADYN calculation with CRD for our flare profiles. We found that while PRD effects are still important, their influence on flare profiles is much smaller than on those of the quiet Sun.

\citet{2015A&A...582A..50K} studied the formation of the 3p-3d level subordinate lines for an M1.8 class flare and found that they usually appear in absorption outside the flaring region but show a significant enhancement during a solar flare, agreeing with our findings. \citet{2015ApJ...806...14P} found that these optically thick lines are formed in the lower chromosphere, and that they appear in emission when there is a large temperature gradient ($\ge$1500 K) throughout a wide range of heights in the chromosphere.

In the following discussion, we will focus only on the \ion{Mg}{2}~k line, as the \ion{Mg}{2}~h line behaves nearly identically. Figure~\ref{mg_loops} shows the temporal evolution of the \ion{Mg}{2}~k intensity resulting from the total contribution of all the threads. The line profile shows two clear emission peaks and a well defined central depression.

At initial times (see panel (a) in Figure~\ref{mg_loops}) the line is symmetric with a reversal in the core. After 20~seconds (panel (b)) the blue peak becomes stronger and the intensity in the core increases, due to strong upflows in the upper chromosphere resulting from the impulsive phase of thread~2. After 60~seconds (Figure~\ref{mg_loops}(e)) the line core show a strong red peak due to downflow velocities and two minor peaks at lower wavelengths in the core associated to upflow velocities of two different threads. During the gradual phase of the flare (Figure~\ref{mg_loops}(h)), the intensity decreases and the profile shows the typical core reversal; plasma motions in the upper chromosphere are responsible of the asymmetries in the line profile.

  \subsubsection{Formation of the \ion{Mg}{2} line emission}
\begin{figure*}[!hbt]
 \centering
 \epsscale{1.05}
  \mbox{
    \hspace{-0.3cm}
    \mbox{\includegraphics[scale=0.34]{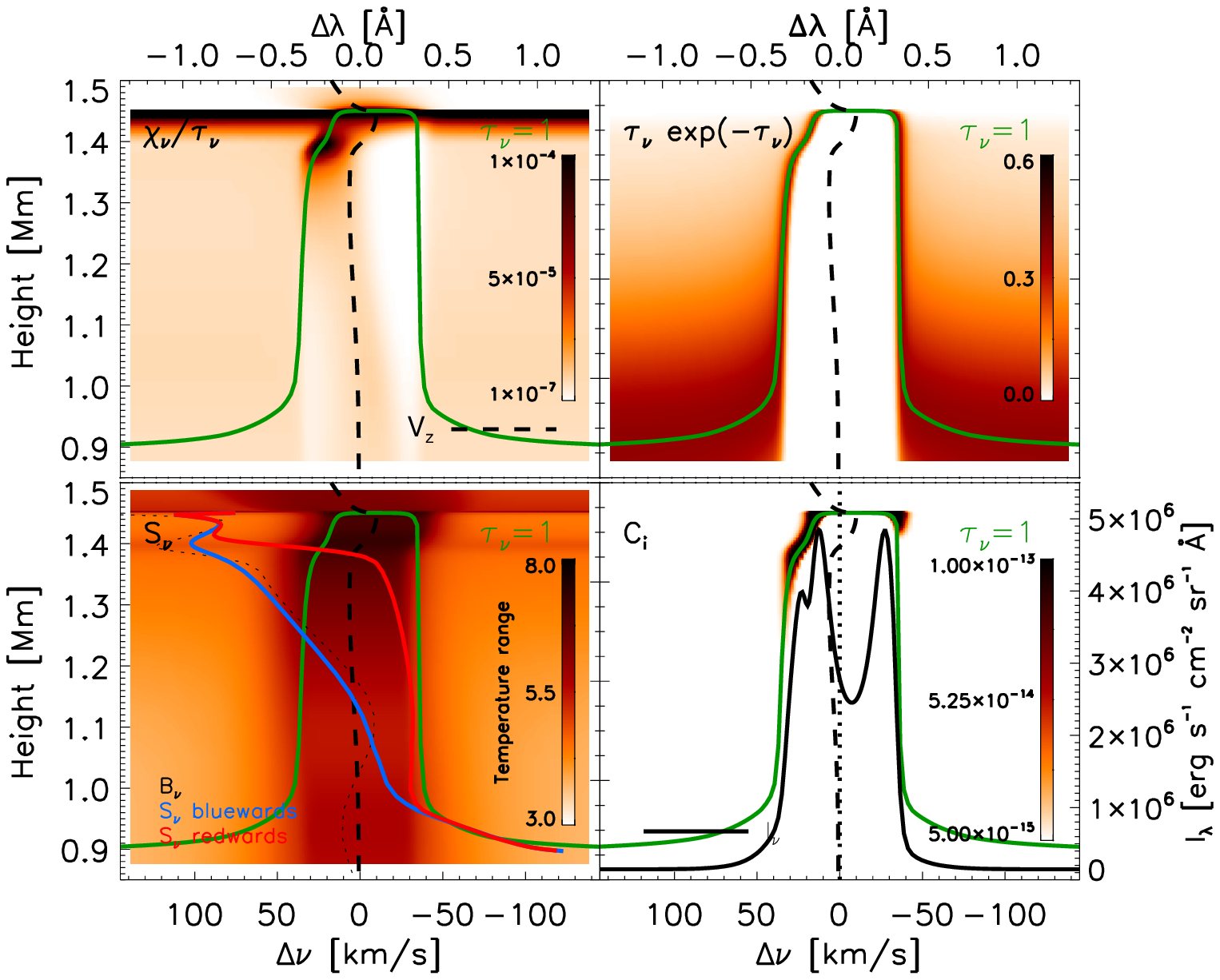}}   
    \hspace{-0.3cm}
    \mbox{\includegraphics[scale=0.34]{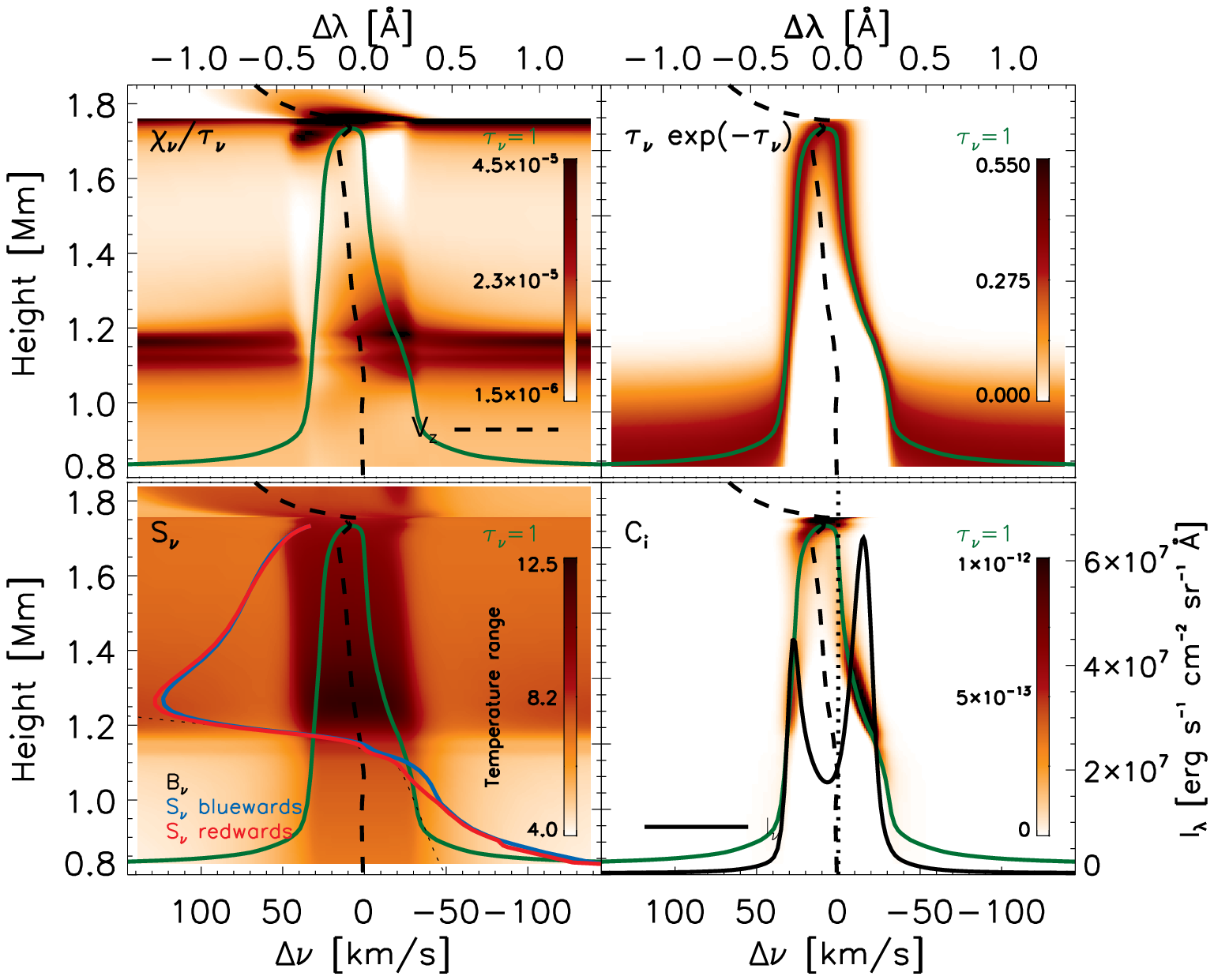}}
    \hspace{-0.3cm}
    \mbox{\includegraphics[scale=0.34]{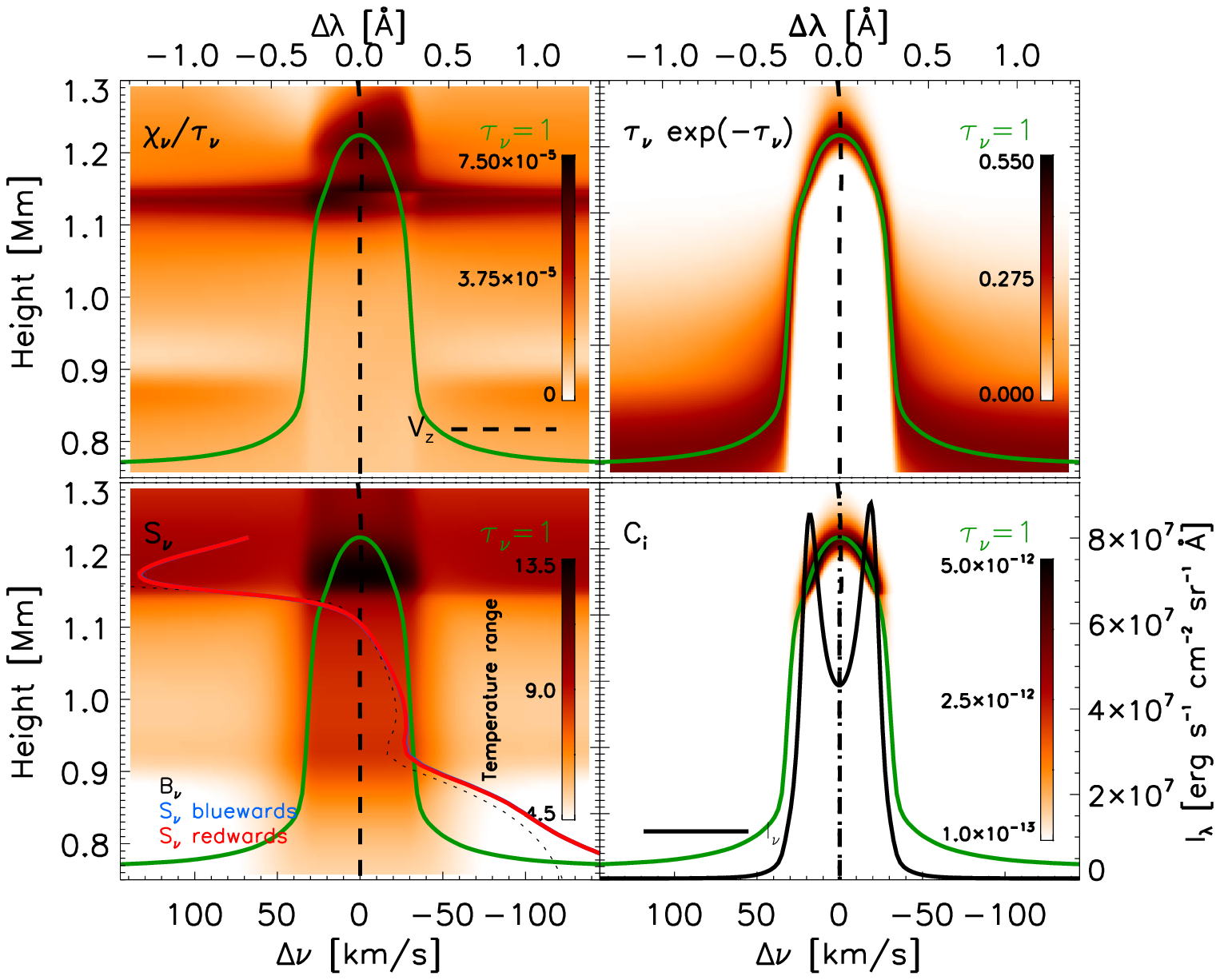}}
   }
    \caption{\ion{Mg}{2}~k intensity contribution function for thread~6 after 38 seconds (left), thread~7 after 19 seconds (middle) and thread~8 after 1 second (right). Each image shows the quantity specified in its top-left corner as a function of frequency (in velocity units, where positive velocities represent upflows and negative velocities, downflows); the top axis is the wavelength from line center and the vertical axis, the height in Mm from the bottom of the loop. Multiplication of the first three panels produces the intensity contribution function in the fourth panel. The $\tau_{\nu}=1$ curve (green solid line) and the vertical velocity (black dashed) are overplotted in each panel. The third panel also shows the Planck function (black dotted line) and the line source function along the $\tau_{\nu}=1$ curve in blue for the part of the $\tau_{\nu}$ curve blueward of its maximum value and red for the part on the red side of the maximum $\tau_{\nu}=1$ height, in temperature units. The lower right panel also contains the emergent intensity profile (solid black line).}
    \label{ci_image_mg_loops6_7_8}
\end{figure*}

To better understand how the line profiles are formed and how they respond to the atmospheric changes, we analyze the formation of the \ion{Mg}{2}~k~ profiles in detail (Figure~\ref{ci_image_mg_loops6_7_8}) using four-panel formation diagrams previously used by \citet{1997ApJ...481..500C}, showing the intensity contribution function and its different components as a function of frequency and height.

We investigate how the atmospheric stratification affects the formation of the line profile and where in the atmosphere the line is formed by separating the formal solution of the transfer equation for emergent intensity in different terms \citep[Equation~\ref{eq_contribution_function}][]{1997ApJ...481..500C}. This decomposition shows that the maximum of the intensity contribution function is located at heights with high source function and at low optical depth, around optical depth unity.

\begin{equation}
I_{\nu}^{0} =
\frac{1}{\mu} \int_{z} S_{\nu} \chi_{\nu} e^{-\frac{\tau_{\nu}}{\mu}} dz = \frac{1}{\mu} \int_{z} C_i \,dz \,\,,
\label{eq_contribution_function}
\end{equation}
where $z$ is the atmospheric height; $\tau_{\nu}$ is the monochromatic optical depth; $\chi_{\nu}$ is the monochromatic opacity per unit volume; $S_{\nu}$ is the source function, defined as the ratio between the emissivity to the opacity of the atmosphere and $C_i$ is the intensity contribution function at height $z$.

The intensity contribution function, $C_i$, is expressed as the product of three terms: $S_{\nu}$, $\frac{\chi_{\nu}}{\tau_{\nu}}$ and $\tau_{\nu}e^{-\frac{\tau_{\nu}}{\mu}}$, whose variations for the relevant threads are shown in Figure~\ref{ci_image_mg_loops6_7_8}.

As shown in Figure~\ref{ci_image_mg_loops6_7_8} that while the line profile resulting from thread~6 is formed within a height range between 0.9~Mm (photons contributing to the wings, at 1~\AA\ from the line center) and 1.46~Mm (photons contributing to the core of the line); most of the \ion{Mg}{2}~k photons of thread~7 are formed within a height range between 0.8 (at the wings, 1~\AA\ from the line center) and 1.73~Mm (in the core of the line); the line profile resulting from thread~8 during its impulsive phase is formed in a narrower height region, between 0.77 (at the wings, 1~\AA\ from the line center) and 1.23~Mm (in the core of the line).

The source function panel in Figure~\ref{ci_image_mg_loops6_7_8} of these three threads clearly demonstrates the frequency-dependence of the line source function in temperature units caused by PRD effects. The blue line represents the source function along the $\tau_{\nu}=1$ curve for the part of the $\tau$ curve blueward of its maximum value; while the red line shows the part on the red side of the maximum $\tau_{\nu}=1$ height. One can see that in thread~6 both lines are decoupled at almost all heights in the chromosphere (between 0.95 and 1.45~Mm, showing the importance of PRD. For thread~7, although the blue and red components of the source function follow the same trend, they are clearly independent. The contribution of thread~8 becomes significant after one second of heating, during its impulsive phase. The red and blue components do not differ, indicating that PRD effects at this time are not as critical as for the other two contributing threads.

The shape of the line profile of thread~6 indicates that, although the line does not show a strong asymmetry in the wings, the core is shifted towards higher wavelengths. This is due to the downflow plasma velocity (negative values) between $\approx$ 1.2 and 1.36~Mm. The core reversal is due to the sudden increase of the source function, changing the direction at 1.46~Mm, where the photons of the line core are formed. This is related to the temperature stratification at this height, shortly below the transition region. The secondary peak at 27 km~s$^{-1}$ is due to a small temperature increase at 1.23~Mm, as shown in Figure~\ref{chromosphere_loops6_7_8}.

The line profile resulting from thread~7 shows a stronger red peak, due to the downflow velocities at 1.64~Mm. 

The third panel of Figure~\ref{ci_image_mg_loops6_7_8} shows a symmetric line profile resulting from thread~8 at the beginning of the impulsive phase. The upflowing velocities shown at 1.62~Mm of Figure~\ref{chromosphere_loops6_7_8}(c) do not affect the shape of the line because the formation height of the core is located in a lower region, at 1.23~Mm.

  \section{Comparison with observations}\label{sect:results}
In general the synthetic line profiles are narrower than the observed ones. We applied a constant microturbulent velocity of 4.5~km~s$^{-1}$ to the H$\alpha$\ and \ion{Ca}{2}~8542~\AA\ lines to simulate the effects of unresolved turbulent motions and a microturbulent velocity of 10~km~s$^{-1}$ to the \ion{Mg}{2} h\&k line profiles.

Observations of \ion{He}{1}~10830~\AA\ were omitted from the comparison because the RADYN code was run without the recent addition of XEUV backwarming from \citet{2015ApJ...809..104A}, which has a significant influence on helium line profiles.

  \subsection{H$\alpha$~and \ion{Ca}{2}~8542~\AA\ Line Profiles}\label{obs_radyn_ha_ca}
We averaged the RADYN model intensities at each wavelength for 3~seconds (the duration of the IBIS line core observations) to compare the synthetic H$\alpha$~and \ion{Ca}{2}~8542 \AA\ profiles with the observations.

\begin{figure*}[thb]
\hspace{-0.2cm}
  \epsscale{2.2}
  \plottwo{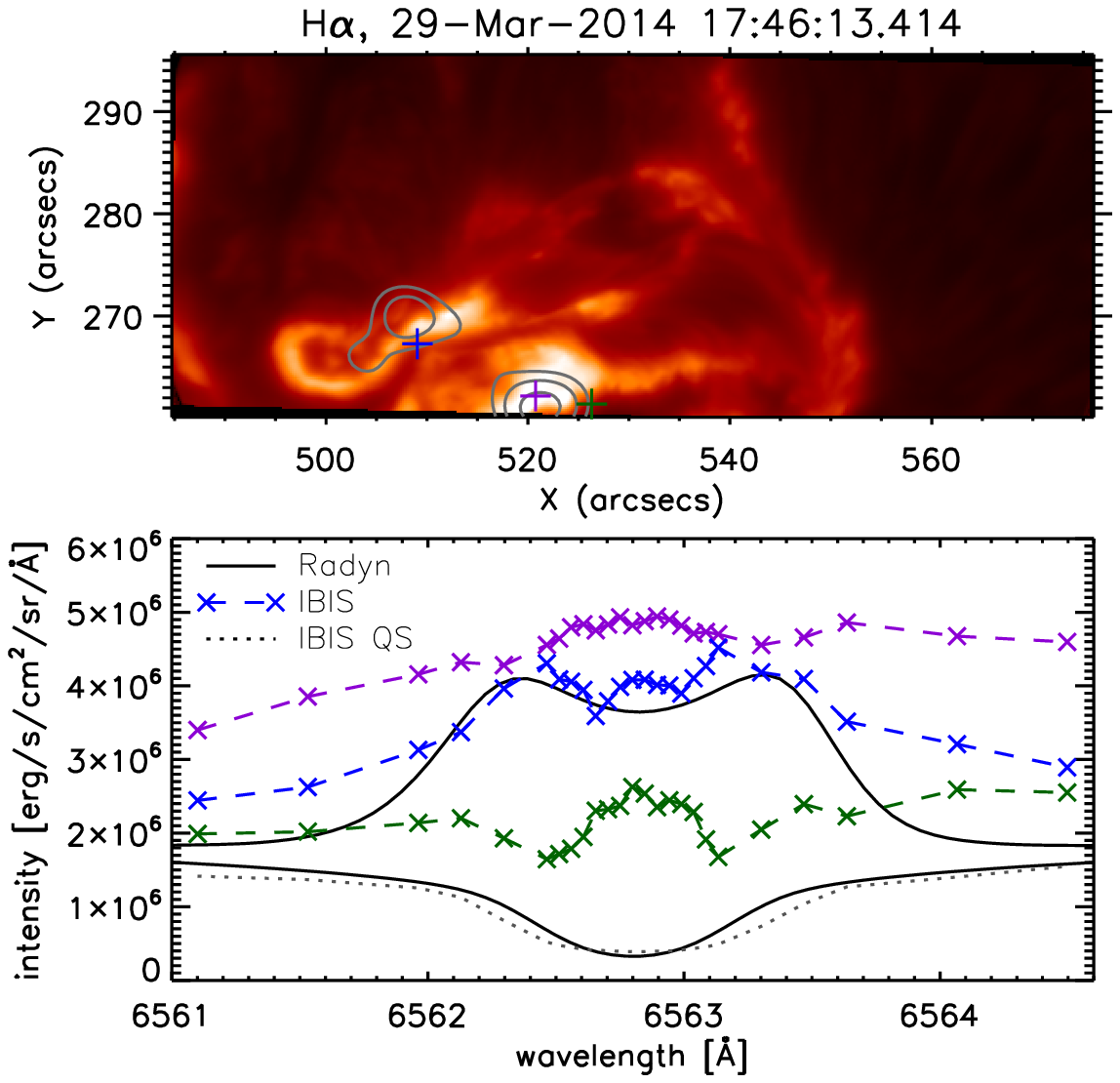}{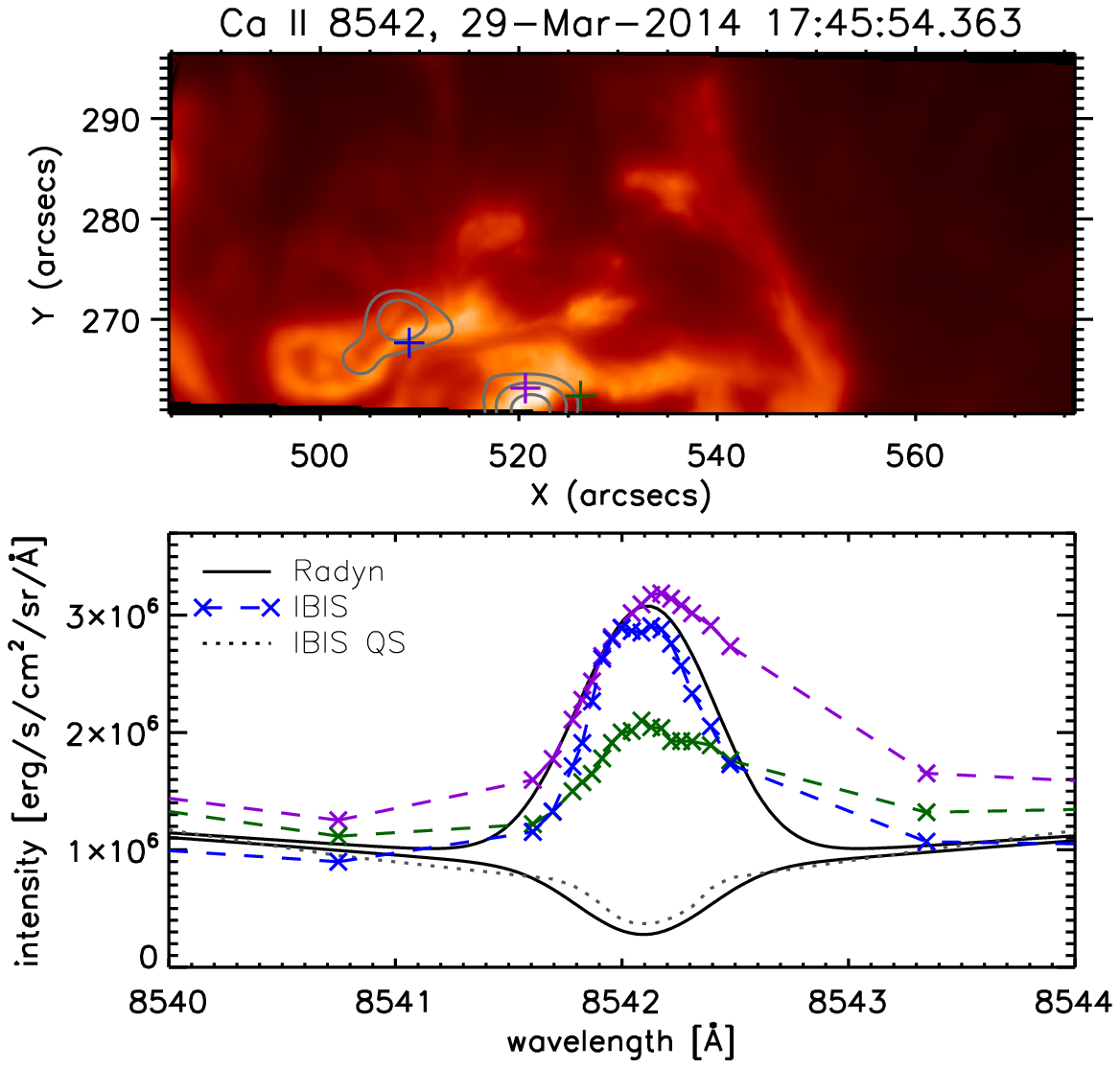}
  \caption{H$\alpha$~and \ion{Ca}{2}~8542~\AA\ line profile for RADYN (black solid line) and IBIS at 17:46:13~UT+18~s and 17:45:54~UT+18~s, respectively, and at different locations along the ribbons of the flare (purple for the southern ribbon and blue for the northern ribbon) and outside (green). The black dotted line is the IBIS quiet Sun profile calibrated to RADYN.}
  \label{ha_radyn_ibis}
\end{figure*}

To properly compare the RADYN profiles with IBIS, we convolved the RADYN quiet Sun profiles with a Gaussian with FWHM of 0.028~\AA\ for H$\alpha$\ and 0.049~\AA\ for \ion{Ca}{2}~8542~\AA, to fit it to the quiet Sun observed profiles. This broadening may be considered as the effect introduced by the microturbulence in the real observations and was applied to all corresponding RADYN data. Finally, to convert the IBIS data into physical units, we matched the IBIS quiet Sun to that of RADYN and applied the conversion factor to the IBIS flare data.

Figure~\ref{ha_radyn_ibis} shows the comparison of IBIS and RADYN for H$\alpha$\ (left panels) and \ion{Ca}{2} 8542 \AA\ (right panels) for one timestep during the impulsive phase (given in the figure). The top row shows intensity images in the respective wavelengths of the line core with {\it RHESSI} contours of [30, 50, 80]\% overplotted in grey. The crosses indicate color-coded locations for which observed profiles are plotted in the bottom row. Each observed profile was averaged over $0\farcs4 \times 0\farcs4$ ($4 \times 4$ pixels). Because of the large variation of the observed profiles, we plot them for three different locations instead of averaging over the whole {\it RHESSI} contour. The averaged contour strongly depends on the selected level and is therefore hard to compare to the simulations. The purple profiles come from the strongest emission region in the southern ribbon/footpoint and their intensity is higher than RADYN's, especially in the line wings. The blue profiles, which come from the northern ribbon/footpoint, agree well with the simulation, both in line intensity and in line width. While the line core seems to have an extra bump in H$\alpha$\ observations, this could be due to an unresolved component or possibly because it took several seconds to scan the whole line profile and the ribbon may evolve faster than this timescale. The observations show larger line wing intensities, which are not reproduced in the simulations. Especially the red wing seems to be higher in observations, indicating missing downflows in the simulations. 

Comparing the temporal evolution of both observed (Figures~\ref{ha_ibis_evol} and \ref{ca_ibis_evol}) and synthetic intensities (Figure~\ref{ha_ca_radyn}), the maximum intensity occurs at the same time, although the intensity resulting from RADYN is $\approx$ a factor of 2 higher in H$\alpha$\ (and slightly less in \ion{Ca}{2}~8542~\AA). While the decay phase seems to be constant, keeping the lines in emission for minutes, in RADYN it occurs faster, concluding that the heating of electrons during the gradual phase of the flares is not enough to keep the chromospheric lines in emission.

\subsection{\ion{Mg}{2} h\&k Line Profiles}\label{obs_radyn_mg}
\begin{figure*}[thb]
\hspace{-0.2cm}
  \epsscale{2.15}
  \plottwo{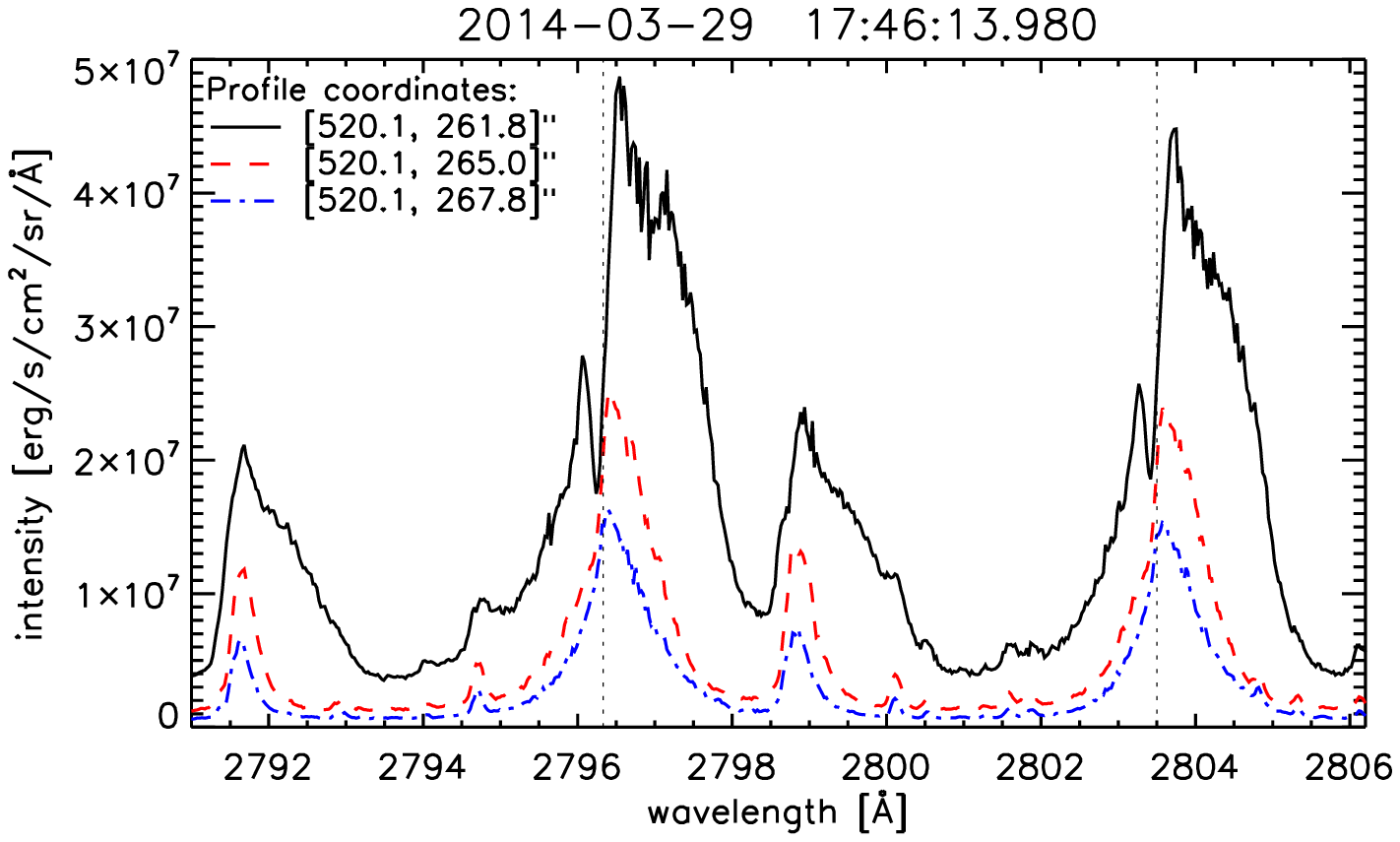}{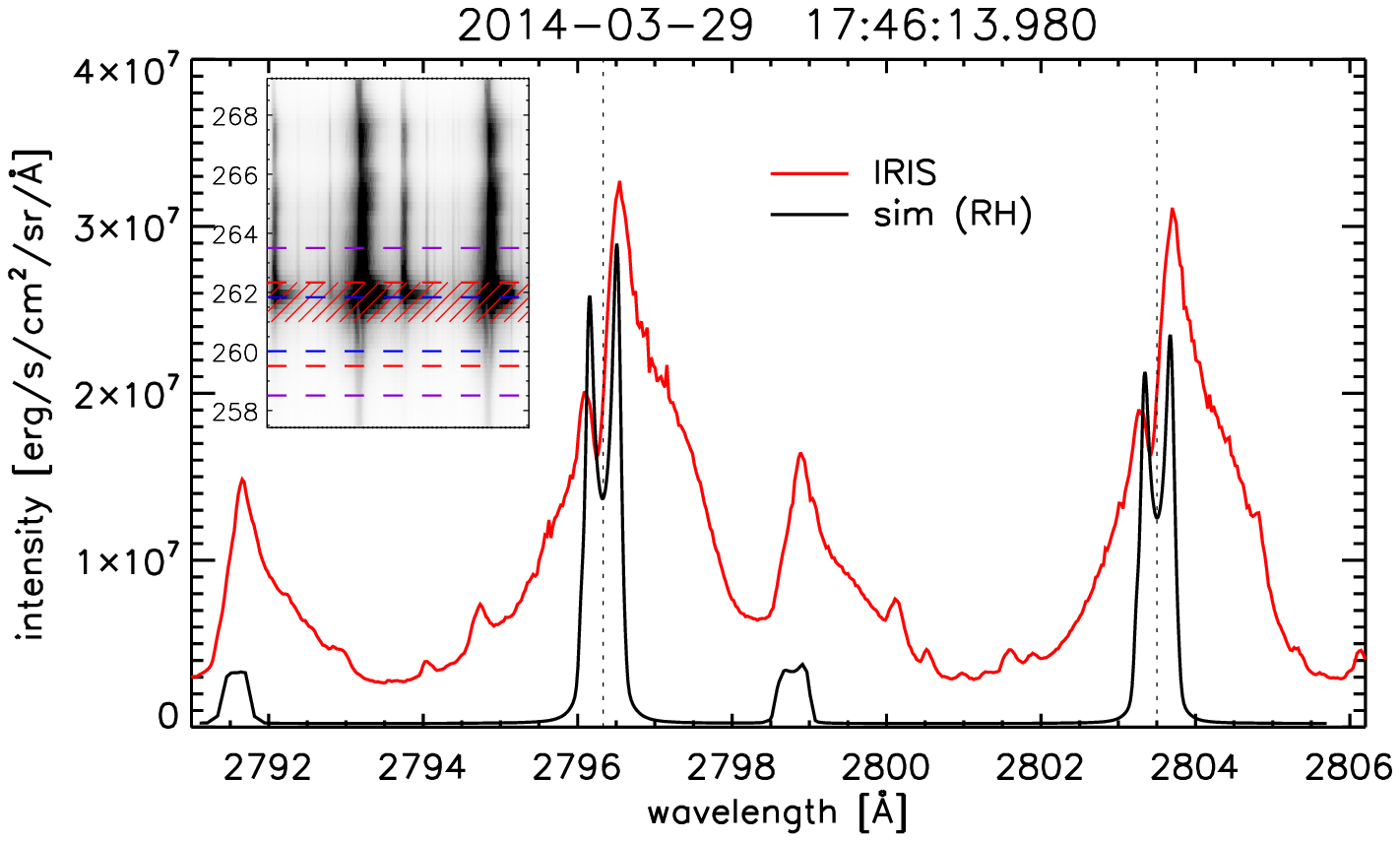}
  \caption{Left: Evolution of the \ion{Mg}{2} h\&k line profiles (black-red-blue). More northern profiles were crossed by the hard X-ray source earlier, indicating they are in a later stage of their evolution due to the flare. Right: Comparison of RH (black) and observations (red). The observations were averaged over half of the 80\% {\it RHESSI} contours, whose location is indicated with red stripes in the inset that shows the Mg spectrum. The lower half of the 80\% contour did not show any flare-related Mg profiles, presumably because it takes a few seconds for the atmosphere to heat, which is why only the upper half of the contour was used for the comparison.}
  \label{mg_rh_iris}
\end{figure*}

The comparison of the \ion{Mg}{2} line profiles is more complex than the above chromospheric lines for two reasons: because the observations covered only the south-western footpoint (see Figure~\ref{iris_slit}) and because the lines are formed in a broad height range in the chromosphere and small differences in the simulated atmosphere below the transition region result in large differences in the core of the line profiles. 

The evolution of the \ion{Mg}{2} line profiles near the hard X-ray footpoints follows a common scheme: At first, the intensity of the line increases abruptly, showing strong downflows i.e.\,enhancements in the red wing, the subordinate \ion{Mg}{2} triplet lines change from absorption into emission. At this time, the central reversal is still visible in the h and k lines. A few seconds later, the intensities decrease and the line profiles show only a single peak, a feature that is not at all common on disk in quiet Sun profiles. The single peak persists for several minutes (in our case for 8 minutes, until the end of the observations), while the intensities are slowly decreasing. This behavior is shown in the left panel of Figure~\ref{mg_rh_iris}. Instead of plotting the temporal evolution from different rasters, which have a relatively slow 75~s cadence, we plot three different spatial positions from the same raster at 17:46:14~UT. Because the hard X-ray source was moving south, these positions can be understood as a temporal evolution after the bombardment of electrons. As can be seen from Figure~\ref{mg_loops}, RADYN does not reproduce the strong downflows, nor the single-peaked profiles.

\begin{figure*}[thb]
\hspace{-0.2cm}
  \epsscale{2.15}
  \plotone{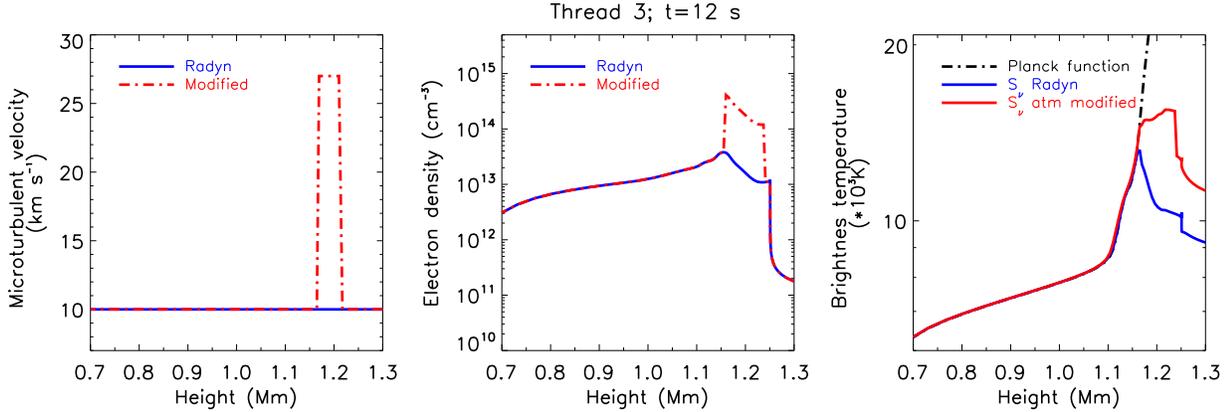}
    \caption{Atmospheric stratification resulting from RADYN for thread~3 at 12~seconds (blue) and the modified one provided to the RH code (red). [Left]: Microturbulent velocity as a function of height; [Middle]: Electron density as a function of height; [Right]: Planck function (black) and source function resulting from the RADYN (blue) and modified (red) atmospheres.}
  \label{atmos_modif}
\end{figure*}

The right panel of Figure~\ref{mg_rh_iris} shows a direct comparison of {\it IRIS} and RH at 17:46:14 UT, during the impulsive phase. The {\it IRIS} data were averaged over half of the 80\% {\it RHESSI} contour. This can be seen from the inset, which shows the Mg spectrum versus the solar latitude. The horizontal lines indicate the northern and southern borders of the [50, 80 ,90]\% {\it RHESSI} contours in the colors purple, red, and blue, respectively. The red shaded area indicates the location used for the average of the {\it IRIS} data. The RH simulations (black) were taken at the same time step for a proper comparison. While the intensities agree relatively well, the line shape and the broadening width obviously do not: our synthetic profile shows a reversal line core and narrow line profiles, while the the observations present a single peak and a broader profile in the line wings.

The \ion{Mg}{2} h\&k lines have very broad wings due to their large opacity in the photosphere \citep{2013ApJ...772...89L}. Moreover 3D effects are very important especially in the line core. In the following section we try a phenomenological procedures to see how the discrepancies can be alleviated. 

  \subsubsection{Testing the discrepancy between the synthetic and observed profiles}
To investigate the reasons for the shape discrepancy, we iteratively modified the RADYN atmosphere and simulated the effects with the RH code. By doing that we aim to explore how the line responds to atmospheric variations and to constrain the atmospheric structure during solar flares. There are four height-dependent parameters that affect the line intensity and shape: temperature, electron density, velocity, and microturbulence. We found that a significantly increased microturbulence is needed to reproduce the broad line wings from the observations; we increased the microturbulent velocity up to 27 km s$^{-1}$ in a narrow region of $\approx$~800~km, lower in the chromosphere, where the line wings are formed (higher microturbulent velocities would distort the line core). This can be understood as highly increased turbulence during flares, which is not included in the 1D RADYN simulation. The formation of this line profile illustrates the intricate formation properties of the \ion{Mg}{2}~h\&k line cores.

We found that the cores of the \ion{Mg}{2} h\&k lines are sensitive to modification of the temperature and electron density in a very narrow region of less than 100~km below the transition region. By increasing the temperature the lines become fainter; while the increase of the electron density in this narrow region increases the intensity of the \ion{Mg}{2} lines. 

In order to get a single peak the electron density has to be increased by almost a factor of 10 shortly below the transition region (see Figure~\ref{atmos_modif}). For a better interpretation of the line formation, the right panel of Figure~\ref{atmos_modif} shows the source and Planck function as function of height for the original line profile with a core reversal (blue) and the line profile showing a single peak core, resulting from the modified atmosphere (red). By increasing the electron density the source function couples to the Planck function, explaining why the core of the line turns into emission.

To match the observed \ion{Mg}{2} emission, the ratio between \ion{Mg}{2}~h and \ion{Mg}{2}~k should be close to one in the model, otherwise the core of the line would be formed in an optically thin environment. The core of the lines is formed in a optically thick region where the densities are so high that the source function does not drop. \citet{2015ApJ...809L..30C} studied how the temperature and density affects these lines in plages and found that by modifying the location of the chromospheric temperature rise towards deeper layers, the core of the \ion{Mg}{2} becomes wider. While in specific cases a change of temperature may affect the core reversal, the density is equally important. Therefore the results of \citet{2015ApJ...809L..30C} may be too simplistic for solar flares.

\begin{figure}[thb]
\hspace{-0.2cm}
  \epsscale{1.05}
  \plotone{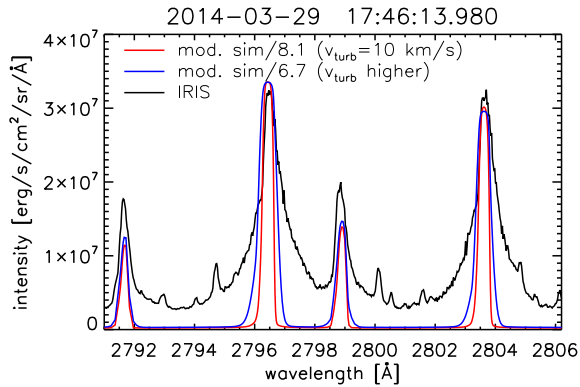}
    \caption{Comparison of the {\it IRIS} \ion{Mg}{2}~h\&k line profiles (black) with those from the RH code, with a modified electron density (red) and micruturbulent velocity (blue) as shown in Figure~\ref{atmos_modif}.}
  \label{mgii_modif}
\end{figure}

In Figure~\ref{mgii_modif} we compare the resulting line profile with the {\it IRIS} observations. While the intensities of the synthetic profiles are higher and the line wings narrower, it nevertheless is the first simulation reproducing the single peaks of Mg, commonly observed during most flares. The higher intensity could be explained as a strong coupling to the local temperature due to high densities.

To account for the whole shape of the line width, one require a line profile consisting of two different gaussians: one with a smaller width similar to that obtained with RH (with a FWHM of $\approx$0.4~\AA) and a second broader one with a FWHM of $\approx$1.3~\AA, corresponding to a velocity higher than 80~km~s$^{-1}$.

Another discrepancy is the emission in between lines: the simulated emission in the wings of the line is much lower than in observations. One possible reason is that RH calculates the opacity of the transitions only in the selected atoms, therefore we may underestimate the opacities.

   \section{Summary and Discussion}\label{sect:conclusions}
In this paper, we have presented a radiative hydrodynamic simulation of an X1.0 class flare and a direct comparison with high-resolution spectroscopic observations in the chromosphere in H$\alpha$~and \ion{Ca}{2}~8542~\AA\ by IBIS and in \ion{Mg}{2} by {\it IRIS}. The temporal variation of the flux of non-thermal electrons was inferred from {\it RHESSI} X-ray spectra and the structure of the initial atmosphere was constrained using the available continuum emission before the flare at four wavelength bands from the UV to the IR.

An important improvement with respect to our previous work, \citet{2015ApJ...804...56R}, is the inclusion of multi-threaded simulations, using the time derivative of the GOES 1--8~\AA\ light curve to estimate the duration of each thread. Multi-thread simulations are more realistic because it is unlikely that only one loop contributes to the observed X-ray emission.

A direct comparison of the line profiles was performed at different times. The synthetic intensity values generally match observations in the chromosphere (H$\alpha$, \ion{Ca}{2}~8542~\AA\ and \ion{Mg}{2}~k). The H$\alpha$, \ion{Ca}{2}~8542~\AA\ synthetic profiles fit in shape to the observations, while the \ion{Mg}{2} line profiles observed by {\it IRIS} are broader in the wings than the synthetic ones. Additionally, flare observations generally show a single peak in Mg, while RADYN simulations (and any other simulations to date) show two peaks, similar to observations outside of flare regions. By iteratively modifying the RADYN atmosphere, we found that we could reproduce the observed single-peaked Mg flare spectra with a strong increase in density below the TR. The too narrow simulated line wings can be interpreted as a lack of microturbulent motions in the simulations at the lower chromosphere, or could also possibly be due to 3D effects, not included in the 1D RADYN and RH codes. By increasing the microturbulence in the upper chromosphere the wings of the simulated \ion{Mg}{2} line profiles fit the observed broad lines better, but we are still missing a broader component to account for the observations.

The \ion{Mg}{2}~h\&k line profiles show a complex formation behavior that spans the entire chromosphere. The variations in temperature, density, and velocity in the chromosphere produce diverse line shapes. The \ion{Mg}{2}~emission is extremely sensitive to changes in a very narrow height range.

The estimated cutoff energy resulting from fitting the {\it RHESSI} spectra is an upper limit, resulting in a lower limit of the flux of electrons. Increasing the electron flux would not help, since it would result in higher intensities in the chromosphere as well and our intensity values are already matching the observations, being slightly higher at certain times.

Our simulation can be improved in the future by including the acceleration and transport of particles: coupling RADYN with the FLARE particle transport and acceleration code \citep{2015ApJ...813..133R} results in stronger downflows, which may influence the asymmetries in the chromospheric line profiles. The inclusion of the improved backwarming component of \citet{2015ApJ...809..104A} would reduce the intensity of the lines.

\acknowledgments
Work performed by F.R.dC., V.P. and W.L. is supported by NASA grants NNX13AF79G and NNX14AG03G. L.K. is supported by a Marie Curie Fellowship. J.C.A is supported by NASA LWS and HSR grants. 
We thank A. Sainz-Dalda, J. Leenaarts, A. Kowalski and F. Effenberger for their helpful discussions.
We gratefully acknowledge the use of supercomputer resources provided by the NASA High-End Computing (HEC) program through the NASA Advanced Supercomputing (NAS) Division at Ames Research Center.
{\it IRIS} is a NASA small explorer mission developed and operated by LMSAL with mission operations executed at NASA Ames Research center and major contributions to downlink communications funded by ESA and the Norwegian Space Centre.

\bibliographystyle{apj}
\bibliography{ms}
\end{document}